\documentclass[letterpaper,twocolumn,10pt]{article}
\usepackage{zhanggroup}

\usepackage{tikz}
\usepackage{amsmath,amssymb,amsfonts}
\usepackage{multirow}
\usepackage{color}
\usepackage{xcolor}
\usepackage{url}
\usepackage{caption}
\usepackage{subcaption}
\usepackage{float}
\usepackage{balance}
\usepackage{graphicx}
\usepackage{textcomp}
\usepackage{booktabs}
\usepackage{colortbl}
\usepackage[ruled,linesnumbered]{algorithm2e}  
\usepackage{algorithmic}
\usepackage{booktabs}
\usepackage{makecell}
\usepackage{tcolorbox}
\usepackage[framemethod=tikz]{mdframed}
\usepackage{pifont}
\newcounter{requirement}
\newcommand{\requirement}[1]{\refstepcounter{requirement}\smallskip\noindent\textbf{Requirement \therequirement.}\label{#1}}

\newcommand{\mypara}[1]{\smallskip\noindent\textbf{#1.}}

\newcommand{\RNum}[1]{\uppercase\expandafter{\romannumeral #1\relax}}
\definecolor{myred}{HTML}{BF0511} 
\definecolor{mygreen}{HTML}{94D05E} 
\definecolor{myblue}{HTML}{305594} 
\newcommand{\attack}{HijackFL\xspace}
\begin{document}
%--------------------------------------------------
\date{}

\title{\bf Model Hijacking Attack in Federated Learning}

\author{
Zheng Li\textsuperscript{1}\ \ \
Siyuan Wu\textsuperscript{2}\ \ \
Ruichuan Chen\textsuperscript{3}\ \ \
Paarijaat Aditya\textsuperscript{3}\ \ \
Istemi Ekin Akkus\textsuperscript{3}
\\
Manohar Vanga\textsuperscript{3}\ \ \
Min Zhang \textsuperscript{2}\ \ \
Hao Li\textsuperscript{2}\ \ \
Yang Zhang\textsuperscript{1}
\\
\\
\textsuperscript{1}\textit{CISPA Helmholtz Center for Information Security} \ \ \ 
\\
\textsuperscript{2}\textit{Institute of Software, Chinese Academy of Sciences} \ \ \
\textsuperscript{3}\textit{Nokia Bell Labs}
}

\maketitle

%--------------------------------------------------
\begin{abstract}
Machine learning (ML), driven by prominent paradigms such as centralized and federated learning, has made significant progress in various critical applications ranging from autonomous driving to face recognition.
However, its remarkable success has been accompanied by various attacks.
Recently, the model hijacking attack has shown that ML models can be hijacked to execute tasks different from their original tasks, which increases both accountability and parasitic computational risks.

Nevertheless, thus far, this attack has only focused on centralized learning.
In this work, we broaden the scope of this attack to the federated learning domain, where multiple clients collaboratively train a global model without sharing their data. 
Specifically, we present \textit{\attack}, the first-of-its-kind hijacking attack against the global model in federated learning. 
The adversary aims to force the global model to perform a different task (called hijacking task) from its original task without the server or benign client noticing.
To accomplish this, unlike existing methods that use data poisoning to modify the target model's parameters, \attack\ searches for pixel-level perturbations based on their local model (without modifications) to align hijacking samples with the original ones in the feature space. 
When performing the hijacking task, the adversary applies these cloaks to the hijacking samples, compelling the global model to identify them as original samples and predict them accordingly.
We conduct extensive experiments on four benchmark datasets and three popular models. 
Empirical results demonstrate that its attack performance outperforms baselines.
We further investigate the factors that affect its performance and discuss possible defenses to mitigate its impact.
\end{abstract}
%--------------------------------------------------

%--------------------------------------------------
\section{Introduction}
%--------------------------------------------------
Machine learning (ML) has progressed rapidly in the past decade. 
Centralized learning and federated learning~\cite{DecentralizedML,KMYRSB16,MMRHA17,NDR20}, as the two most representative paradigms, have played a crucial role in this advancement. 
In centralized learning, ML models are trained on a central server where the training data are centrally gathered.
In contrast, federated learning enables collaborative training across multiple decentralized devices, ensuring local privacy and data ownership. 
These approaches have led to unprecedented achievements of ML in various applications, such as image classification~\cite{TWKTI20,HCLWMW18,TPLFB15,ABCGL22}, healthcare diagnostics~\cite{KEEKF15,SWFJH10, BFDB11}, and authentication systems~\cite{ZDH17,KSMB16}.

Despite being popular, ML models are vulnerable to a range of security and privacy attacks. 
For instance, the adversary can craft a malicious dataset to train the target model for various malicious purposes.
Such attacks are commonly known as training time attacks, i.e., attacks that interfere with the training process of the target model.
Backdoor attacks~\cite{GDG17,SWBMZ22,LMALZWZ18,BVHES20} and poisoning attacks~\cite{BNL12,JOBLNL18,STLLXCS18} are two of the most popular training time attacks. 
In backdoor attacks, the adversary manipulates the target model to produce malicious output when facing any input that contains specific triggers, while behaving normally on clean data.
In contrast, poisoning attacks try to jeopardize the model's utility performance on clean data.
Both attacks have been demonstrated across centralized learning and federated learning.

\begin{figure}[t]
    \centering
    \includegraphics[width=0.9\linewidth]{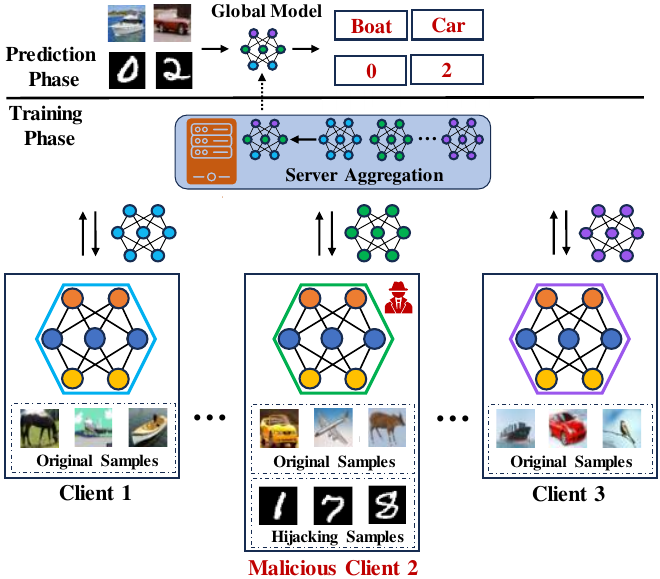}
    \caption{The overview of model hijacking attack in federated learning. The original task is CIFAR-10, while the hijacking task is MNIST.}
    \label{fig:overview}
\end{figure}

In this paper, we focus on a new type of attack, the model hijacking attack, where the adversary aims to force the target model to perform a completely different task, known as the hijacking task.
For instance, the target model is designed for the CIFAR-10 task, while an adversary aims to repurpose it for a different task, such as MNIST.
Successful model hijacking attacks can cause severe consequences.
The first is about accountability, in which the model owner can unknowingly be held responsible for unintended and potentially illegal or unethical services.
Another significant risk is parasitic computing, where the model owner pays for model maintenance and training while the adversary utilizes it for free or for its own purposes.

In NDSS'22, Salem et al.~\cite{SBZ22} presented the first model hijacking attack against machine learning models, or more precisely, centralized learning models.
As the data provider, the adversary can manipulate the target model's training dataset by integrating the hijacking dataset into the original one, i.e., the data poisoning approach.
Consequently, models trained on such a modified dataset will have their parameters altered, leading to a change in their behavior. 
Si et al.~\cite{SBZS23} also utilize the data poisoning approach to hijack NLP models.
However, the data poisoning approach used in both Salem et al.~\cite{SBZ22} and Si et al.~\cite{SBZS23} is only applicable to centralized learning models.

\mypara{Motivation} In this study, we consider the model hijacking attack in the federated learning paradigm.
We exclusively focus on federated learning in the image domain.
As depicted in \autoref{fig:overview}, the adversary, acting as a malicious client, attempts to repurpose the global model intended for the original task (e.g., CIFAR 10) to perform the hijacking task (e.g., MNIST).
Intuitively, two unique characteristics of federated learning hinder existing model hijacking attacks~\cite {SBZ22,SBZS23}.
First, federated learning enables multiple participants to train a global model collaboratively.
This collaborative characteristic implies that a malicious client may only participate in a few training rounds to submit their modified model parameters to the server. 
In most training rounds, only benign local model parameters are submitted to the global model. 
This results in a high probability of the hijacking effect being diminished or eliminated.
Second, modified local model parameters submitted by the adversary may differ from benign model ones, which can be detected by the central server.
Our experiments (see \autoref{sec:attack_performance})  confirm the aforementioned: the adversary cannot implant the hijacking in the FL domain using the data poisoning approach.
To overcome these limitations, we make the first attempt to answer, ``Is it possible to hijack a cooperatively trained global model without relying on any modifications to it?''

\mypara{Our Contributions} In this paper, we introduce the first model hijacking attack against the global model in the FL domain, called \attack.
Specifically, \attack\ achieves the goal of hijacking by adding pixel-level perturbations, referred to as ``cloaks,'' into the hijacking samples. 
Using the local model as a foundation, the adversary optimizes these cloaks to ensure that the hijacking samples closely resemble the original samples within the feature space.
Once the cloaks are optimized, the local model can identify hijacking samples with added cloaks as original samples, thus predicting them to the original classes according to the adversary's predetermined class mapping.
Furthermore, due to the high similarity between the local and global models, these cloaks will exhibit a high degree of transferability, thus allowing the global model to accurately classify hijacking samples with added cloaks.
Since the adversary does not modify their local model but only submits benign local model parameters to the server, \attack successfully addresses the challenges faced by existing attack methods.

We conduct extensive evaluations using four benchmark computer vision datasets and three widely used models. 
Empirical results demonstrate that \attack has no side effect in the utility of the global model and, furthermore, outperforms existing attacks~\cite{SBZ22,SBZS23}.
For instance, when the global model is ResNet-18, and the original dataset is CIFAR-10, \attack\ can achieve an attack success rate of 92.75\% for the MNIST hijacking dataset. 
In contrast, the existing attacks only achieve an attack success rate of around 10\%.
Furthermore, we conduct a comprehensive ablation study to analyze the factors that may affect the attack performance and also explore two possible defenses to mitigate our attack.

Abstractly, our contributions can be summarized as follows:
\begin{itemize}
    \item We propose the first-of-its-kind model hijacking attack against the global model in federated learning, called \attack.
    \item Without modifying the local model, \attack\ adds cloaks to the hijacking samples so that these samples have features highly similar to the original samples, thus allowing the global model to classify them as the original classes according to the adversary's predetermined class mapping.
    \item We conduct extensive evaluations, and empirical results show that our attacks achieve significant performance regarding both utility and attack success rate. We further explore the various factors that may affect the attack performance, as well as two potential defenses.
\end{itemize}

%--------------------------------------------------
\section{Preliminaries}
%--------------------------------------------------
We first describe how federated learning works.
Then, we present the model hijacking attack in centralized learning.

%--------------------------------------------------
\subsection{Federated Learning}
%--------------------------------------------------
Federated Learning (FL)~\cite{DecentralizedML,KMYRSB16,MMRHA17,NDR20} is a machine learning paradigm that allows multiple clients to collaboratively train a model without sharing their training data.
Basically, FL can be divided into two kinds, depending on how the training data is split among clients. 
The most common setting is known as horizontal FL, with the training data instances being split across the sample space. 
That is, each client hosts different training data instances sharing the same features. 
Another FL setting is vertical FL, where each client hosts the same training data instances but owns different and non-overlapping features.
In this work, we only consider the most common setting -- horizontal FL (briefly called FL throughout this paper).

Formally, considering an FL training process involves $n$ clients, each client holds a set of private training data.
At each training round $t$, the central server selects $m$ clients and sends them the current global model $G^t$.
Each selected client, e.g., $i$th of $m$ clients, updates this model to a new local model $F^{t+1}_{i}$ by training on its private data, and sends the difference $F^{t+1}_{i} - G^t$ back to the central server.
The central server then aggregates the received model parameters from these $m$ clients, e.g., averaging these parameters (FedSGD~\cite{SWMS19}), to obtain the new global model $G^{t+1}$:
\begin{equation}\label{eq_1}
G^{t+1}=G^t+\frac{\eta}{n} \sum_{i=1}^m\left(F_i^{t+1}-G^t\right)
\end{equation}
Here, the $\eta$ is a global learning rate set by the central server, controlling the fraction of the global model that is updated.
The central server and all parameters share the same model architecture, and only the model parameters $F^{t+1}_{i} - G^t$ are communicated regularly between them.
Normally, the training does not stop until the global model converges, a malicious client thus always has an opportunity to share model updates with the central server.

%------------------------------------------------
\subsection{Model Hijacking Attack}\label{naive_attack}
%------------------------------------------------
Model hijacking attack is a new class of training time attacks against ML models.
In centralized learning, Salem et al.~\cite{SBZ22} assume the adversary acts as a data provider and follows a similar implementation as data poisoning attacks, i.e., poisoning the training dataset to modify the target model's parameters.
Two attack methods are proposed in Salem et al.~\cite{SBZ22}.
The first is to directly mix the original dataset with the hijacking dataset.
This achieves the upper bound on attack performance yet causes a drawback: it can be detected by the model owner as the original and hijacking datasets are visually different.
Thus, Salem et al. introduce an extra model to transform the hijacking dataset into a format visually similar to the original dataset.
Subsequently, Si et al.~\cite{SBZS23} presented an attack similar to the second one mentioned above, but targeting NLP models.
Both studies demonstrate the applicability of the model hijacking attack to the centralized learning domain.

\mypara{Note} It is important to note that model hijacking differs from data poisoning and backdooring.
Data poisoning~\cite{BNL12,JOBLNL18,STLLXCS18} aims to \textit{jeopardize the models’ utility}, and backdooring~\cite{GDG17,LMALZWZ18} links \textit{the trigger inserted in the original sample} (e.g., a white square at the corner) to specific model output.
Though the backdoor attack can be considered a specific instance of model hijacking to some extent, hijacking a model is to execute \textit{an entirely different task}, regardless of the original one. 
This implies that achieving the hijacking objective is more challenging.
Further, the adversary is free to determine the hijacking task in the hijacking attack.

%------------------------------------------------
\subsection{Threat Model in FL}\label{sec:threatmodel}
%------------------------------------------------
We envision the adversary as a malicious client in the FL training process. 
The adversary's goal is the same as the model hijacking attack in centralized learning ~\cite{SBZ22,SBZS23}, where all benign clients expect the global model to perform the original task (e.g., CIFAR-10), while the adversary expects the global model to perform a hijacking task (e.g., MNIST).

We assume that the adversary, like other benign clients, does not break the FL training protocol.
The adversary only exchanges model parameters between itself and the central server without making any additional assumptions about the central server or benign clients.
Furthermore, we assume that the adversary holds an additional dataset for the hijacking task, which is a basic and necessary assumption for all current model hijacking attacks.
We emphasize that the adversary has full control over its own local training process, i.e., performs any operation on its local model.

A successful model hijacking attack in the FL domain should satisfy the following two requirements.
We refer to the deployed global model (after training and attack) as the hijacked global model.

\requirement{} \label{req1} \textit{The hijacked global model should have a similar performance as the clean global model on its original dataset.}

\requirement{} \label{req2} \textit{The hijacked global model should correctly classify most samples in the hijacking dataset.}

%------------------------------------------------
\section{Methodology}
%------------------------------------------------
To hijack the global model in FL, we introduce the \attack\ attack.
We first present the overview of \attack.
Then, we present the detailed design of \attack.

\begin{figure}[!t]
\centering
\begin{subfigure}{0.49\columnwidth}
\includegraphics[width=\columnwidth]{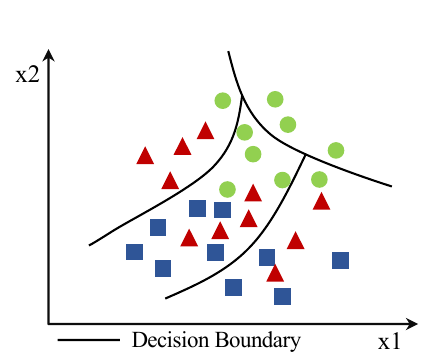}
\caption{Without Cloak}
\label{fig:intuition_a}
\end{subfigure}
\begin{subfigure}{0.49\columnwidth}
\includegraphics[width=\columnwidth]{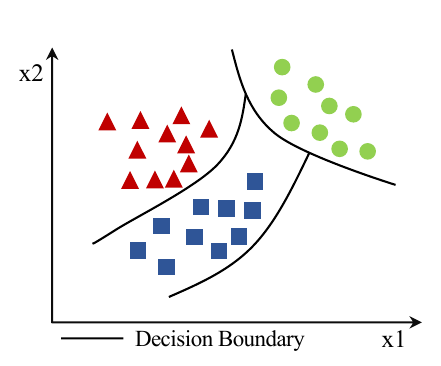}
\caption{With Cloak}
\label{fig:intuition_b}
\end{subfigure}

\caption{The overview of cloaking the hijacking samples, which can then be classified by the global model. The background is the decision boundary of the original task. The colored points represent the features of hijacking samples.}
\label{fig:intuition}
\end{figure}

%--------------------------------------------------
\subsection{Overview}
%--------------------------------------------------
ML models are trained to extract and identify features from input samples, which they employ for prediction. However, their feature identification can be misled by adversarial example attacks~\cite{PMSW18,PMGJCS17,PMJFCS16}, where slight perturbations to input samples alter the model's feature perception. 
Our attack exploits this vulnerability to mislead the global model's perception of features extracted from hijacking samples toward the original samples.

During FL training, the adversary updates their local models by fetching the global model from the central server. 
Leveraging the local model, the adversary then learns slight perturbations (called cloaks) for the hijacking samples, enabling the local model to extract features close to those from the original samples. 
If the global model remains relatively stable in subsequent training rounds, the cloaks learned on the local model can be transferred to the global model.
This transferability enables the global model to also extract features from the hijacking samples that are similar to the original samples.
The global model thinks it works correctly, as it perceives the features of input samples within the ``original dataset.''
Through this process, the adversary successfully accomplishes the goal of hijacking the global model.

We illustrate this in \autoref{fig:intuition} using a simplified 2D visualization of the feature space. 
As depicted in \autoref{fig:intuition_a}, the decision boundary represents the local/global model trained on the original dataset, but it fails to distinctly classify the hijacking samples into separate regions. 
However, after adding the cloaks, the hijacking samples will become clearly separable into different regions (\autoref{fig:intuition_b}), enabling accurate predictions by the model. 
We emphasize that the adversary does not modify the local model, but only the hijacking sample by adding cloaks.
Thus, the adversary exchanges only clean model parameters with the server, just like a benign client. 
This prevents the two challenges of modifying local model parameters via poisoning training datasets: (1) being detected by the server and (2) the hijacking effect being removed.

%--------------------------------------------------
\subsection{Computing Cloak Perturbations}
%--------------------------------------------------
But how do we determine what perturbations (we call them ``Cloaks'') to apply to the hijacking samples?
An effective cloak could cause the global model to associate the hijacking samples with features similar to the original samples.
Intuitively, the closer the features extracted from the hijacking samples are to the original samples, the higher the probability that the global model will accurately identify the hijacking samples.

In the following, we describe the core component of \attack\ for computing cloaks for hijacking samples, with the goal of making their features extracted by the global model highly similar to those extracted from original samples.

\begin{figure}[!t]
\centering
\begin{subfigure}{0.49\columnwidth}
\includegraphics[width=\columnwidth]{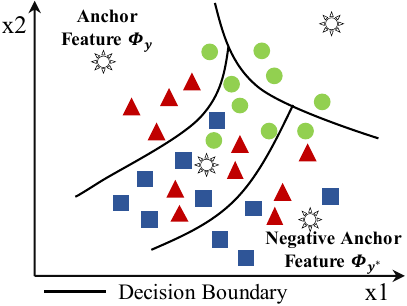}
\caption{Without Cloak}
\label{fig:class_specific_cloak_a}
\end{subfigure}
\begin{subfigure}{0.49\columnwidth}
\includegraphics[width=\columnwidth]{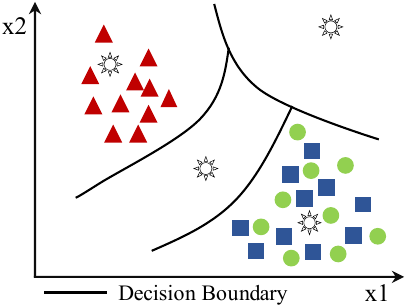}
\caption{With Cloak for \textcolor{myred}{\scalebox{1.2}{$\blacktriangle$}} }
\end{subfigure}

\begin{subfigure}{0.49\columnwidth}
\includegraphics[width=\columnwidth]{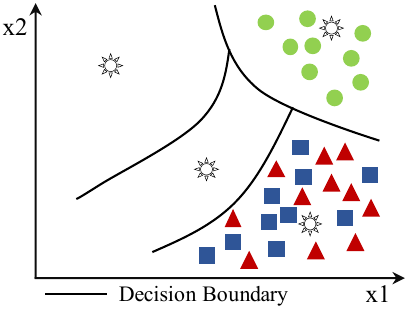}
\caption{With Cloak for \textcolor{mygreen}{\raisebox{-.3ex}{\scalebox{1.5}{$\bullet$}}}} 
\end{subfigure}
\begin{subfigure}{0.49\columnwidth}
\includegraphics[width=\columnwidth]{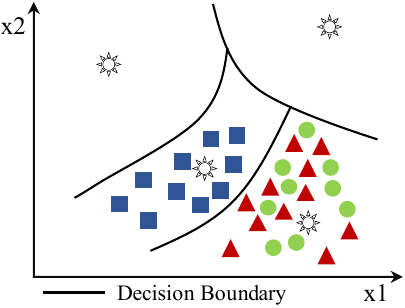}
\caption{With Cloak for \textcolor{myblue}{\rule{1.3ex}{1.3ex}}}
\end{subfigure}

\caption{The overview of the class-specific cloaking. The background is the decision boundary of the original task. The colored points represent the features of hijacking samples. The negative anchor feature $\Phi_{y^{\ast}}$ is fixed in the lower right region.}
\label{fig:class_specific_cloak}
\end{figure}

\mypara{Note} Our discussion will use the following notations/definitions:
\begin{itemize}
    \item $y$: A class of the original dataset (original class).
    \item $h$: A class of the hijacking dataset (hijacking class).
    \item $x_{h}$: A hijacking sample of class $h$.
    \item $\delta$: A cloak perturbation for a hijacking sample.
    \item $x_h \oplus \delta$: Cloaked version of $x_h$.
    \item $\bar{h}$: A hijacking class beyond $y$.
    \item $\mathcal{X}_h$: All hijacking samples of class $h$.
    \item $\Phi$: Feature extractor (the local model $F$ excluding the final classification layer).
    \item $\Phi(x_h)$: A feature vector extracted from $x_h$.
    \item $\Phi_y$: A feature vector that predicts the original class $y$ with close to 100\% confidence, called \underline{anchor feature} of $y$ (see \autoref{fig:class_specific_cloak_a}). 
\end{itemize}

\mypara{Cloaking to Minimize Feature Deviation}
Formally, given a hijacking sample $x_h$ and a original class $y$, our ideal design is to find a cloak perturbation $\delta$ that minimizes the deviation between the sample's feature $\Phi(x_h\oplus\delta)$ and the anchor feature $\Phi_y$:

\begin{equation}
\begin{gathered}
\min _\delta \operatorname{\textit{Dist}}\left(\Phi\left(x_h \oplus \delta\right), \Phi_y\right), \\
\end{gathered}
\end{equation}
Where $\operatorname{\textit{Dist}}$ calculates the distance of two feature vectors.
Once the adversary succeeds in finding a cloak perturbation $\delta$ and adds it to the hijacking sample $x_h$, the cloaked sample $x_h \oplus \delta$ will be identified and extracted with anchor feature $\Phi_y$, which the model then predicts as the original class $y$.

\mypara{Class-specific Cloaking}
When generating cloaks for a hijacking task, the adversary will create class-specific cloaks based on their held hijacking dataset, i.e., $\delta$ is class-dependent. 
Specifically, depending on the held hijacking dataset, the adversary optimizes the cloak for a particular hijacking class; therefore, samples within this class can be extracted to the same or similar features.
This ensures that the cloak exhibits a high degree of generalizability, allowing it to map new, unseen real-world hijacked samples from that class to similar features, thus predicting them to the same original class.
Therefore, instead of producing individual cloaks for each hijacking sample, we propose creating a single, generalizable cloak for each hijacking class.
For each hijacking class $h$, all samples $\mathcal{X}_h$ share a common cloak $\delta_h$.
To achieve this, the adversary will pair any hijacking sample $x_h \in \mathcal{X}_h$ with an anchor feature $\Phi_y$. 
The optimization problem is formulated as follows:

\begin{equation}\label{equ_3}
\begin{gathered}
\min _{{\delta}_h} \operatorname{\textit{Dist}}\left(\Phi\left(x_h \oplus \delta_h\right), \Phi_y\right), \\
\text { subject to } \forall x_h \in \mathcal{X}_h.\\
\end{gathered}
\end{equation}
This class-specific cloak optimization starts with $\delta=0$ (no perturbation) and iteratively updates $\delta_h$ on all hijacking samples $\mathcal{X}_h$.
The adversary then finds a cloak perturbation $\delta_h$ and adds it to any hijacking sample $x_h \in \mathcal{X}_h$. 
The cloaked samples will be predicted to the original class $y$.
In this way, the adversary establishes a class mapping $h \Rightarrow y$.
To establish another class mapping, the adversary performs the above optimization again and finally searches for a new cloak.

The above implementation, however, tends to search for overlearned cloaks so that the model only responds to cloaks and ignores the hijacking samples they cloak.
In other words, given an unseen hijacking sample for query, adding a different cloak to it will be predicted as a different original class.
This poses a challenge to the adversary to choose a suitable cloak for it.
To address this, we introduce the negative anchor feature $\Phi_{y^{\ast}}$ (\autoref{fig:class_specific_cloak_a}), which is fixed for any hijacking samples. 
As illustrated in \autoref{fig:class_specific_cloak}, the cloak $\delta_h$ not only establishes a class mapping $h \Rightarrow y$, but also maps all hijacking samples beyond class $h$ to the original class $y^{\ast}$, i.e., $\bar{h} \Rightarrow y^{\ast}$. 
Now, given an unseen hijacking sample $x$, it can be accurately classified as the original class $y$ only when added with its associated cloak. 
However, if added with any other cloaks, it will \textit{always} be classified as the original class $y^{\ast}$. 
This design enables the adversary to easily determine which cloak to add to a given query sample, as only one cloak will not result in the specific original class $y^{\ast}$.

To this end, in addition to \autoref{equ_3}, we further pair any hijacking samples beyond class $h$, i.e., $x_{\bar{h}} \in \mathcal{X}_{\bar{h}}$, with the negative anchor feature $\Phi_{y^{\ast}}$ and minimize the distance between $\Phi({x_{\bar{h}}\oplus\delta_{h}})$ and $\Phi_{y^{\ast}}$:
\begin{equation}\label{eq_4}
\begin{gathered}
\min _{\delta_h} \left(\operatorname{\textit{Dist}}\left(\Phi\left(x_h \oplus \delta_h\right), \Phi_y\right) + \lambda \operatorname{\textit{Dist}}\left(\Phi\left({{x}_{\bar{h}}} \oplus \delta_h\right), \Phi_{y^{\ast}}\right)\right), \\
\text { subject to } \forall x_h \in \mathcal{X}_h \text{ and } \forall {x_{\bar{h}}}\in \mathcal{X}_{\bar{h}}.\\
\end{gathered}
\end{equation}
The coefficient $\lambda$ balances the two $\operatorname{\textit{Dist}}$, and is set to 1.2 in our implementation.
The second $\operatorname{\textit{Dist}}$ actually serves as a penalty for the cloak perturbation, to reduce overlearning. 

%--------------------------------------------------
\subsection{\attack\ Pipeline}
%--------------------------------------------------
After introducing the core component of \attack, we now present the entire attack pipeline.
The attack pipeline can be divided into four stages: Class Mapping, Anchor Features Computing, Cloaks Computing, and Executing.
In the following section, we consider a simple case where \textit{the adversary's hijacking dataset comprises three classes ($h\in {0, 1, 2}$), and the global model's original dataset comprises four classes ($y\in {0, 1, 2, 3}$).}

\mypara{Class Mapping}
During this stage, the adversary needs to predefine a class mapping between the hijacking classes $h$ and the original classes $y$.
To this end, we define a hard-coded mapping function $\textit{{M}}(\cdot)$ that maps a class from $h$ to $y$.
In the above case, $\textit{{M}}(\cdot)$ may be defined to assign the first 3 original classes in a one-to-one mapping to the 3 hijacking classes, as adopted by Salem et al.~\cite{SBZ22}.

Nevertheless, it's plausible that hijacked samples from certain classes may exhibit a stronger inclination to be associated with a particular original class, rather than being equally inclined towards all original classes (verified in Section \ref{sec:impact_class_mapping}). 
Motivated by this, we propose a greedy-based class mapping approach aimed at facilitating the optimization of cloaks.

\begin{figure}[t]
    \centering
    \includegraphics[width=0.9\linewidth]{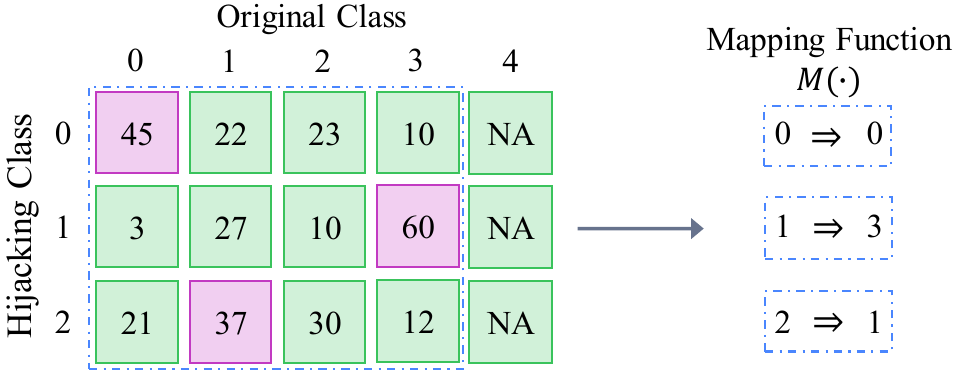}
    \caption{The greedy-based class mapping. The number of hijacking samples is the same for each hijacking class (i.e., each row). Each cell's value indicates the count of hijacking samples from that hijacking class mapped to the original class. NA indicates this original class is used for negative anchor features and not assigned to a particular hijacking class.}
    \label{fig:greedy_class_mapping}
\end{figure}

Specifically, we feed equal-sized hijacking samples from each class to the local model and record the frequency of their mapping to each original class. 
See an illustration in \autoref{fig:greedy_class_mapping}. 
Notably, we exclude the last original class, which serves as $y^{\ast}$ for negative anchor feature (used later), and focus solely on the remaining original classes. 
We then identify the hijacking class that maps to the largest number of original classes, e.g., $1 \Rightarrow 3$ shown in \autoref{fig:greedy_class_mapping}. 
Subsequently, we exclude the selected hijacking and original classes, repeating this process with the remaining classes until all hijacking classes are assigned. 
This results in sequences such as $0 \Rightarrow 0$ and ultimately $2 \Rightarrow 1$. 
Thus, the class mapping function $\textit{M}(\cdot)$ is established as $[0\Rightarrow 0, 1\Rightarrow 3, 2\Rightarrow 1]$. 
Further details of the greedy-based class mapping can be found in Appendix \autoref{fig:relabel_details}.

\mypara{Anchor Features Computing}
After determining the class mapping $\textit{M}(\cdot)= [0\Rightarrow 0, 1\Rightarrow 3, 2\Rightarrow 1]$ and negative original class $y^{\ast}=4$.
The adversary is currently preparing anchor features that significantly contribute to predicting each assigned original class.
To do this, for the assigned original class $y$, we initialize a random sample $r$ from Gaussian noise and feed it the local model $F$ to obtain a probability $F_y(r)$ of its class $y$.
We then maximize the probability by optimizing the random sample as follows:
\begin{equation}\label{anchorfeature}
\begin{gathered}
\min _r -\text{log} F_y(r)\\
\end{gathered}
\end{equation}
We optimize this loss with Adam and terminate until the probability is over a threshold.
After we obtain an optimized sample, we feed it into the local model and extract its features from the intermediate layer, i.e., the anchor feature $\Phi_y$ we aim to obtain.
Note that the threshold should be set very close to 1; only then can we claim that the searched anchor feature indeed significantly contributes to the prediction of class $y$.
In our implementation, we set the threshold to 0.99.
In this way, the adversary access the anchor features $\Phi_{y\in[0,1,2,3]}$: 3 anchor features and 1 negative anchor feature.

\begin{figure}[t]
    \centering
    \includegraphics[width=0.9\linewidth]{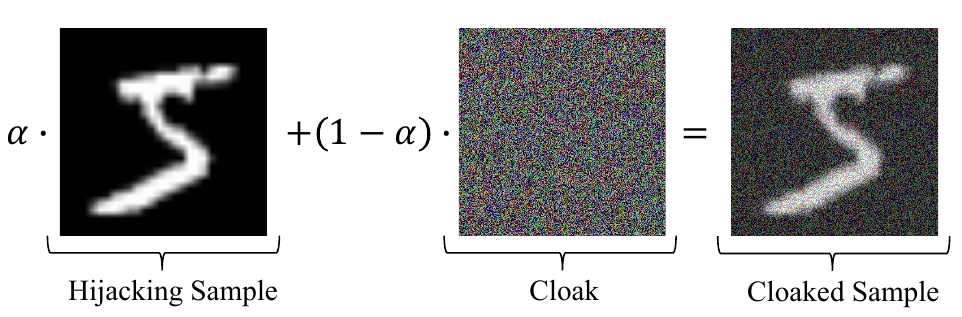}
    \caption{The convex combination of the hijacking sample and the cloak.
    }
    \label{fig:addition}
\end{figure}

\mypara{Cloaks Computing}
For each hijacking class $y$, we initialize a cloak $\delta_h$ from Gaussian noise and transform it using a sigmoid function $\sigma$ to ensure its values align with the hijacking samples, i.e., $255 \times \sigma(\delta_h) \in [0, 255]$.
We then gather all hijacking samples of this class and take a convex combination as follows:
\begin{equation}
\begin{gathered}
{x_h}\oplus \delta_h = \alpha{x_h} +(1-\alpha )  \delta_h \\
\text { subject to } \forall x_h \in \mathcal{X}_h.\\
\end{gathered}
\end{equation}
We use a parameter $\alpha \in [0, 1]$ to balance the weight/influence of $x_h$ and $\delta_h$ on cloaked samples.
\autoref{fig:addition} present an illustration.
In our implementation, we set $\alpha$ to 0.5.
We also study the effect of $\alpha$ by varying it in \autoref{sec:alpha}.

\begin{algorithm}  
\caption{Cloaks Computing.}
\label{alg:cloaks}  
\KwIn{Feature extractor $\Phi$; Anchor features $\Phi_{y\in[0,1,2,3]}$; Hijacking samples of different classes $\mathcal{X}_{h \in [0,1,2]}$; Class mapping $\textit{{M}}(\cdot)=[0\Rightarrow 0, 1\Rightarrow 3, 2\Rightarrow 1]$; Iteration $T$; Hyperparameter $\lambda$. }
\KwOut{ A list \textbf{Cloaks};} 
initialize an empty list: \textbf{Cloaks} = []\;
set $y^{\ast}= 3$ \;
select negative anchor feature: $\Phi_{y^{\ast}}$\;
\For{$h \in [0,1,2]$}
{   initialize cloak $\delta_h$\; 
    select anchor feature $\Phi_{\textit{{M}}(h)}$\;
\For{ $t  \leftarrow  0$ \text{to} $T$}
{  
    select ${x_h}$ from $\mathcal{X}_{h}$\;
    select negative samples $x_{\bar{h}}$ from  $\mathcal{X}_{\bar{h}}$\;
 
    $loss_1$ = $\operatorname{\textit{Dist}}\left(\Phi\left(x_h \oplus \delta_h\right), \Phi_{\textit{{M}}(y)}\right)$\;
    
    $loss_2$ = $\operatorname{\textit{Dist}}\left(\Phi\left(x_{\bar{h}}\oplus \delta_h\right), \Phi_{y^{\ast}}\right)$\;

    $\delta_h$ = $argmin(loss_1 + \lambda loss_2)$
}
add $\delta_h$ to \textbf{Cloaks}\;
}
return \textbf{Cloaks};  
\end{algorithm} 

The adversary then randomly selects equal-size hijacking samples beyond class $h$, i.e., $x_{\bar{h}}$, and obtains their cloaked versions by adding the same cloak $\delta_h$. 
We then feed both $x_h\oplus \delta_h$ and $x_{\bar{h}}\oplus \delta_h$ to the local model's feature extractor $\Phi$, and minimize the two feature distances, i.e., $\operatorname{\textit{Dist}}\left(\Phi\left(x_h \oplus \delta_h\right), \Phi_y\right)$ and $\operatorname{\textit{Dist}}\left(\Phi\left(x_{\bar{h}} \oplus \delta_h\right), \Phi_{y^{\ast}}\right)$.
We use L2 distance as the loss function in feature space, $\operatorname{\textit{Dist}}(\cdot)$.
The L2 distance has good mathematical properties, i.e., convexity and gradient continuity, which can help optimize convergence (see discussion in the Appendix \autoref{sec:convergence}).
Lastly, to ensure the cloaked sample pixels remain in the correct range ($[0, 255]$), we transform the optimized cloak values into $[0, 255]$ by $255 \times \sigma(\delta_h) $.
\autoref{alg:cloaks} presents the core algorithm of cloaks computing based on the aforementioned case.

\begin{figure}[t]
    \centering
    \includegraphics[width=0.95\linewidth]{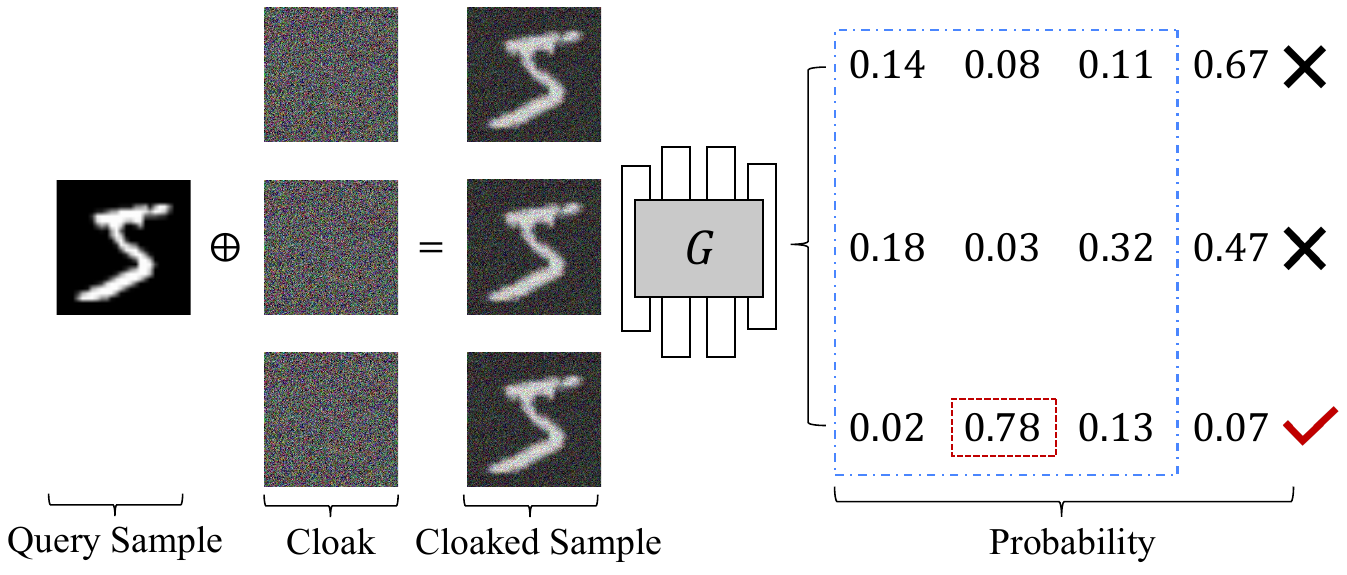}
    \caption{The final execution of \attack against a remotely deployed global model.
    }
    \label{fig:execution}
\end{figure}

\mypara{Executing}
Based on the held hijacking dataset, the adversary has created 3 cloaks for the class mapping $\textit{{M}}(\cdot)=[0\Rightarrow 0, 1\Rightarrow 3, 2\Rightarrow 1]$ 
The adversary can now execute the attack on the remotely deployed global model.
Given a new hijacking sample, the adversary cannot directly determine which cloak should be added to this sample.
Instead, as shown in \autoref{fig:execution}, the adversary obtains 3 cloaked samples by adding 3 cloaks to the given sample respectively.
The adversary then queries the global mode for each cloaked sample to obtain its 4-dimensional probability.
The adversary observes only the first 3 dimensions of the probability, and selects the class with the highest probability as the final prediction, 
Finally, the adversary can determine its hijacking class by reversing the class mapping ${\textit{M}^{-}}(\cdot)=[0 \Rightarrow 0, 3 \Rightarrow 1, 1 \Rightarrow 2]$.

%--------------------------------------------------
\section{Experimental Setup}
%--------------------------------------------------

%--------------------------------------------------
\subsection{Datasets}\label{setup}
%--------------------------------------------------
To evaluate \attack, we employ four well-established computer vision benchmark datasets: MNIST~\cite{MNIST}, CIFAR-10~\cite{CIFAR},  GTSRB~\cite{GTSRB}, and TinyImageNet-100~\cite{TinyImageNet}. 
Here's a brief introduction to each dataset:

\mypara{MNIST} MNIST is a gray image dataset consisting of handwritten digits. It contains 60,000 training images and 10,000 test images, with each image representing a single digit from 0 to 9.

\mypara{CIFAR-10} CIFAR-10 is a dataset comprising 60,000 color images divided into 10 classes. Each image in CIFAR-10 belongs to one of the following categories: airplane, automobile, bird, cat, deer, dog, frog, horse, ship, or truck.
 
\mypara{GTSRB} GTSRB is an image collection consisting of 43 traffic signs. 
It consists of over 51,839 color images, with 39,209 used for training and 12,630 used for testing. 
Due to the high imbalance of each class, we select the top ten classes with the highest number of samples to create a more balanced dataset.

\mypara{TinyImageNet-100} TinyImageNet-100 is a subset of the larger ImageNet dataset, containing 100 object classes with 500 training images and 50 validation images per class.

We use CIFAR-10 and TinyImageNet-100 as the original datasets, while MNIST and GTSRB are the hijacking datasets.
The different combinations of these datasets and dataset sizes are presented in \autoref{tab:task_combination_train} and \autoref{tab:task_combination_test}.
Since the last original class should be fixed for negative pairs, when the original dataset is CIFAR-10, we randomly sample 9 classes from MNIST and GTSRB as the hijacking dataset.
This is actually an inherent limitation of the model hijacking attack, i.e., the original dataset should have more class labels than the hijacking dataset.
We will discuss this further in \autoref{sec:limitation}.
Lastly, following Salem et al.~\cite{SBZ22}, we rescale all datasets to the image size of $32\times 32$, and convert the grayscale MNIST to three-channel images by repeating the same values in all channels.

%--------------------------------------------------
\subsection{Model Structures}
%--------------------------------------------------
As aforementioned, we focus on the machine learning classification setting in this work. 
Thus, following Salem et al.~\cite{SBZ22}, we also use the popular image classification model as our target model, namely ResNet-18~\cite{HZRS16}, MobileNet-V2~\cite{SHZZC18}, and ShuffleNet-V2~\cite{MZZS18}.
Each of these model architectures enjoys its own unique property and has been applied in a wide range of applications.
Besides, all clients and the central server share the same model architecture.

%--------------------------------------------------
\subsection{Federated Learning Settings}
%--------------------------------------------------
We detail the concrete settings from two aspects: global settings and local settings.

\mypara{Global Settings} Following prior work~\cite{BVHES20}, we set the number of training rounds to 200, and the total number of clients contributing to the global model is $n=50$. 
In each round, the central server selects $m=5$ clients, and each client is chosen with equal probability, ensuring that all 50 clients have been involved in one round after 10 training rounds. 
Furthermore, the central server averages the received updates from the selected 5 clients in each training round.
The global learning rate is set to $\eta = 10$.

\mypara{Local Settings} Each client controls a same-sized subset of the full original dataset, e.g., one client controls 2000 samples of the TinyImageNet-100 dataset locally.
In each round, all the selected clients train the local model using the same training settings, i.e., an optimizer of SGD~\cite{B10}, a learning rate of 0.1, and a local training epoch of 2.

\begin{table}[!t]
    \centering
    \caption{Different combinations for the original and hijacking datasets used in \textit{FL training phase}.}
    \scalebox{0.70}
    {
    \begin{tabular}{l|r||l|r||l}
    \toprule
    \multirow{2}{*}{TaskSet} & \multirow{2}{*}{Original Dataset} & \multirow{2}{*}{Hijacking Dataset} &\multicolumn{2}{c}{\multirow{2}{*}{Dataset Size}} \\
     &  &  &\multicolumn{2}{c}{}  \\
     \midrule
     \RNum{1} & CIFAR-10 (10)       & MNIST (9)      & 60000 & 54000       \\
     \midrule 
     \RNum{2} & CIFAR-10 (10)       & GTSRB (9)       & 60000 & 15120\\
     \midrule
     \RNum{3} & TinyImageNet-100 (100) & MNIST (10)       & 50000 & 60000 \\
     \midrule
     \RNum{4} & TinyImageNet-100 (100) & GTSRB (10)      & 50000 & 16110 \\
     \bottomrule
    \end{tabular}
    }
    \label{tab:task_combination_train}
\end{table}

\begin{table}[!t]
    \centering
    \caption{The size of original and hijacking datasets used in \textit{FL prediction/testing phase}.}
    \scalebox{0.7}
    {
    \begin{tabular}{l|r||l|r||l}
    \toprule
    \multirow{2}{*}{TaskSet} & \multirow{2}{*}{Original Dataset} & \multirow{2}{*}{Hijacking Dataset} &\multicolumn{2}{c}{\multirow{2}{*}{Dataset Size}} \\
     &  &  &\multicolumn{2}{c}{}  \\
     \midrule
     \RNum{1} & CIFAR-10 (10)       & MNIST (9)      & 10000 & 9000        \\
     \midrule 
     \RNum{2} & CIFAR-10 (10)       & GTSRB (9)       & 10000 & 4320  \\
     \midrule
     \RNum{3} & TinyImageNet-100 (100) & MNIST (10)       & 10000 & 10000 \\
     \midrule
     \RNum{4} & TinyImageNet-100 (100) & GTSRB (10)      & 10000 & 4800  \\
     \bottomrule
    \end{tabular}
    }
    \label{tab:task_combination_test}
\end{table}

%--------------------------------------------------
\subsection{Hijacking Attack Settings}\label{HAS}
%--------------------------------------------------
Throughout all the FL training rounds, the adversary consistently plays the role of a benign client, submitting benign local model updates trained exclusively on the original dataset to the central server. 
Additionally, we configure the adversary to mount cloak computation at the $150$th training round, referred to as the \underline{hijacking round}. 
The optimization of anchor features involves 500 iterations on a random sample with a learning rate of 0.005. 
Similarly, cloak computation optimization involves 100 iterations with a learning rate of 0.005. 
These settings are determined based on observing convergence around these iterations during the optimization process.

%--------------------------------------------------
\subsection{Baseline Attacks}
%--------------------------------------------------
Existing model hijacking attacks~\cite{SBZ22,SBZS23} aim to hijack centralized ML models to perform different tasks. 
The core idea of existing work is to utilize data poisoning to achieve the goal of hijacking. 
We generalize this data poisoning approach to the federated learning setting as a baseline attack. 
In addition, we consider adapting the data poisoning approach to model poisoning based on the idea of backdooring the FL model proposed by Bagdasaryan et al.~\cite{BVHES20}.

\mypara{Data Poisoning} 
The adversary can poison the training dataset, achieving the goal of hijacking when the model is trained on this poisoned data. 
Salem et al.~\cite{SBZ22} and Si et al.~\cite{SBZS23} employ this method. 
Specifically, Salem et al.~\cite{SBZ22} introduce two attack methods. 
The first involves mixing the original dataset with the hijacking dataset and then using this poisoned dataset to train the target model.
This achieves the upper bound on attack performance yet causes a drawback: it can be easily detected by the model owner since samples in the original and the hijacking datasets can be significantly different. 
Thus, Salem et al.~\cite{SBZ22} introduce an additional model to transform the hijacking dataset into one that is visually similar to the original dataset, which is then mixed with the original dataset to train the target model.
Si et al.~\cite{SBZS23} adopt a similar approach by using an additional model to transform the hijacking dataset.
In this study, we apply both of these methods to the FL scenario, called \underline{Data Poison (naive)} and \underline{Data Poison (transform)}.
We emphasize that both methods are essentially data poisoning attacks aimed at altering the target model's parameters, while our attack avoids such modifications. 

\mypara{Model Poisoning} Model poisoning (also called model replacement) is first proposed in backdooring FL models by Bagdasaryan et al.~\cite{BVHES20}.
This approach exploits the fact that federated learning gives malicious clients direct influence over the global model, enabling significantly more powerful attacks than data poisoning. 
Specially, the adversary can substitute the new global model $G_{t+1}$ with a malicious model $X$ in \autoref{eq_1}:
\begin{equation}
X=G^t+\frac{\eta}{n} \sum_{i=1}^m\left(F_i^{t+1}-G^t\right)
\end{equation}
As the global model converges, these deviations start to cancel out, i.e., $\sum_{i=1}^{m-1}\left(F_i^{t+1}-G^t\right)\approx 0$.
Thus, the adversary can submit the model update as follows:
\begin{equation}
    F_i^{t+1}=\frac n\eta X-(\frac n\eta-1)G^t-\sum_{i=1}^{m-1}(F_i^{t+1}-G^t)\approx\frac n\eta(X-G^t)+G^t
\end{equation}
This adversary scales up the weights of the malicious model $X$ by $\gamma=\frac n\eta $ to ensure that the backdoor survives the averaging and the global model is replaced by $X$.
In this work, we adapt the two aforementioned data poisoning methods to model poisoning, called \underline{Model Poison (naive)} and \underline{Model Poison (transform)}.
Specifically, we scale up the adversary's submitted model parameters to survive the averaging.
We make a strong assumption for the adversary that they know the exact FL settings, i.e., $n=50$ and $\eta=10$.
Thus,  we set the scale-up parameter $\gamma=\frac n\eta = 5 $.

\begin{table*}[!ht]
\definecolor{mygray}{gray}{0.9}
\centering
\caption{The utility of the clean global models and the hijacked global models on the clean test set.}
\scalebox{0.7}
{
\begin{tabular}{l|c|c|c|c|c|c|c}
\toprule
Model&\multirow{2}{*}{TaskSet} & Clean & \multicolumn{5}{c}{\cellcolor[HTML]{E6E6E6} Hijacked Global Model} \\
Architecture&&Model&\cellcolor[HTML]{E6E6E6}Data Poison (naive) &\cellcolor[HTML]{E6E6E6}Data Poison (transform)&\cellcolor[HTML]{E6E6E6}Model Poison (naive)&\cellcolor[HTML]{E6E6E6}Model Poison (transform)  &\cellcolor[HTML]{E6E6E6}Ours \\
\midrule
\multirow{4}{*}{ResNet-18}&\RNum{1}& 0.8803 & \cellcolor[HTML]{E6E6E6}0.8822 & \cellcolor[HTML]{E6E6E6}0.8639 & \cellcolor[HTML]{E6E6E6} 0.8639 & \cellcolor[HTML]{E6E6E6} 0.8774 & \cellcolor[HTML]{E6E6E6}0.8803 \\
&\RNum{2}& 0.8803 & \cellcolor[HTML]{E6E6E6}0.8803& \cellcolor[HTML]{E6E6E6}0.8713& \cellcolor[HTML]{E6E6E6} 0.8809 & \cellcolor[HTML]{E6E6E6} 0.8483 & \cellcolor[HTML]{E6E6E6}0.8803 \\
&\RNum{3}& 0.5460 & \cellcolor[HTML]{E6E6E6}0.5362 & \cellcolor[HTML]{E6E6E6}0.5390 & \cellcolor[HTML]{E6E6E6} 0.5456 & \cellcolor[HTML]{E6E6E6} 0.3806 & \cellcolor[HTML]{E6E6E6}0.5460  \\
&\RNum{4}& 0.5460 & \cellcolor[HTML]{E6E6E6}0.5344 & \cellcolor[HTML]{E6E6E6}0.5328 & \cellcolor[HTML]{E6E6E6} 0.5436 & \cellcolor[HTML]{E6E6E6} 0.3102 & \cellcolor[HTML]{E6E6E6}0.5460 \\
\midrule
\multirow{4}{*}{MobileNet-V2}&\RNum{1}& 0.8592 & \cellcolor[HTML]{E6E6E6}0.8459 & \cellcolor[HTML]{E6E6E6}0.8041 & \cellcolor[HTML]{E6E6E6} 0.8554 & \cellcolor[HTML]{E6E6E6} 0.7466& \cellcolor[HTML]{E6E6E6}0.8592 \\
&\RNum{2}& 0.8592 & \cellcolor[HTML]{E6E6E6}0.8597 & \cellcolor[HTML]{E6E6E6}0.8247 & \cellcolor[HTML]{E6E6E6} 0.8529 & \cellcolor[HTML]{E6E6E6} 0.8121& \cellcolor[HTML]{E6E6E6}0.8592 \\
&\RNum{3}& 0.4472 & \cellcolor[HTML]{E6E6E6}0.4592 & \cellcolor[HTML]{E6E6E6}0.3884& \cellcolor[HTML]{E6E6E6} 0.4350 & \cellcolor[HTML]{E6E6E6} 0.2994 & \cellcolor[HTML]{E6E6E6}0.4472  \\
&\RNum{4}& 0.4472 & \cellcolor[HTML]{E6E6E6}0.4598 & \cellcolor[HTML]{E6E6E6}0.4218& \cellcolor[HTML]{E6E6E6} 0.4564 & \cellcolor[HTML]{E6E6E6} 0.2564 & \cellcolor[HTML]{E6E6E6}0.4472 \\
\midrule
\multirow{4}{*}{ShuffleNet-V2}&\RNum{1}& 0.8705 & \cellcolor[HTML]{E6E6E6}0.8716 & \cellcolor[HTML]{E6E6E6}0.8489 & \cellcolor[HTML]{E6E6E6} 0.8674 & \cellcolor[HTML]{E6E6E6} 0.8174& \cellcolor[HTML]{E6E6E6}0.8746 \\
&\RNum{2}& 0.8746 & \cellcolor[HTML]{E6E6E6}0.8726 & \cellcolor[HTML]{E6E6E6}0.8765 & \cellcolor[HTML]{E6E6E6} 0.8730 & \cellcolor[HTML]{E6E6E6} 0.8438 & \cellcolor[HTML]{E6E6E6}0.8746 \\
&\RNum{3}& 0.5178 & \cellcolor[HTML]{E6E6E6}0.5188 & \cellcolor[HTML]{E6E6E6}0.4822 & \cellcolor[HTML]{E6E6E6} 0.5150 & \cellcolor[HTML]{E6E6E6} 0.3334 & \cellcolor[HTML]{E6E6E6}0.5178  \\
&\RNum{4}& 0.5178 & \cellcolor[HTML]{E6E6E6}0.5132 & \cellcolor[HTML]{E6E6E6}0.5028 & \cellcolor[HTML]{E6E6E6} 0.5252 & \cellcolor[HTML]{E6E6E6} 0.3752 & \cellcolor[HTML]{E6E6E6}0.5178 \\
\bottomrule
\end{tabular}}
\label{table:utility}
\end{table*}

%--------------------------------------------------
\subsection{Evaluation Metrics}
%--------------------------------------------------
To evaluate the performance of our attack, we adopt two metrics, namely utility and attack success rate, following prior works~\cite{SBZ22,SBZS23}.

\mypara{Utility}
Utility measures how closely the performance of the hijacked global model matches that of a clean (non-hijacked) global model on the original testing dataset (\autoref{req1}). 
A higher similarity in performance indicates a more successful attack.
To evaluate the utility, we first train a clean global model by initializing the federated learning training process using only the original training dataset. 
After the training ends, we assess the performance of both the hijacked and clean global models using a original testing dataset.
This original testing dataset is sourced from the same distribution as the original training dataset.

\begin{figure*}[!t]
\centering
\begin{subfigure}{0.59\columnwidth}
\includegraphics[width=\columnwidth]{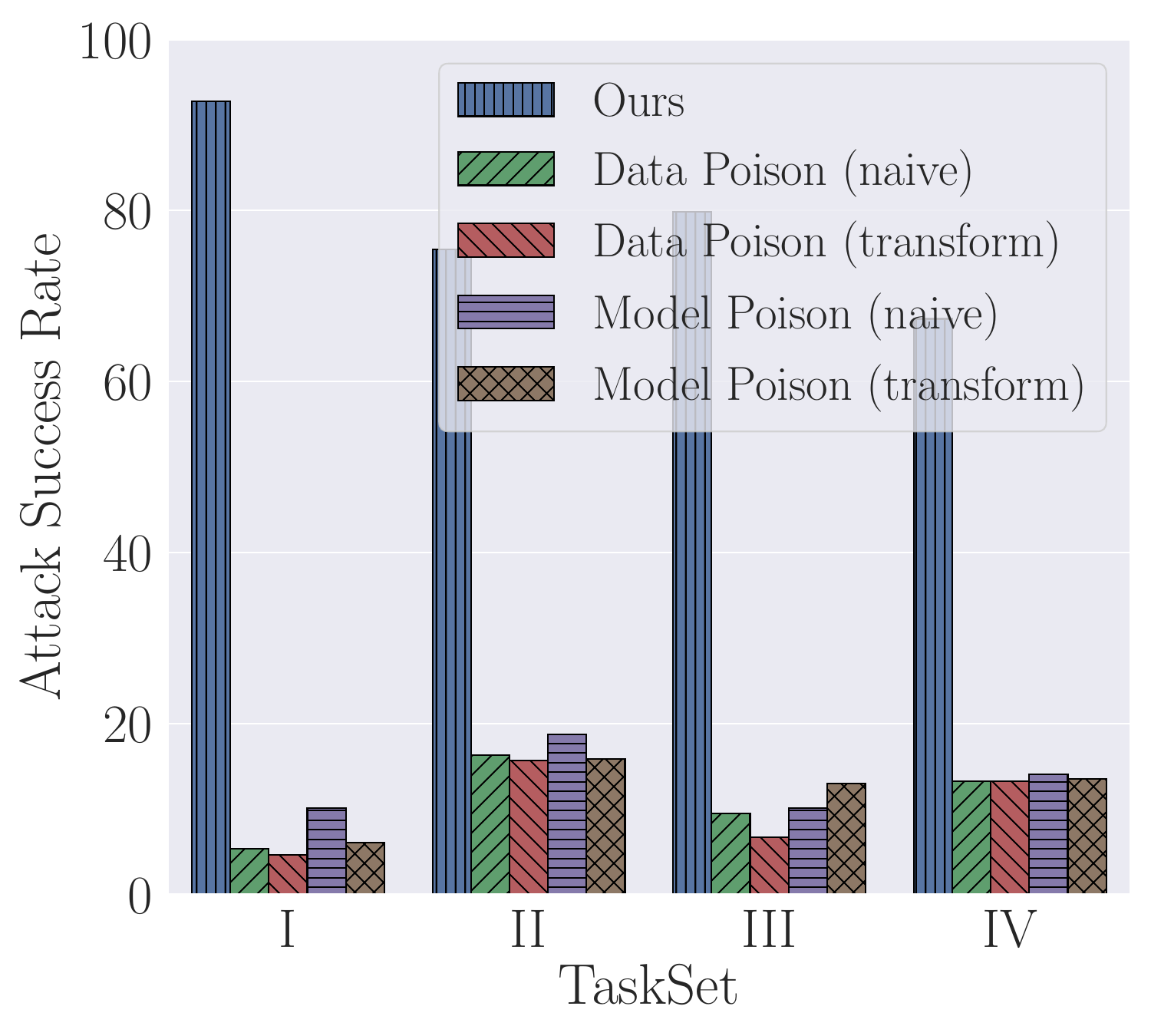}
\caption{ResNet-18}
\end{subfigure}
\begin{subfigure}{0.59\columnwidth}
\includegraphics[width=\columnwidth]{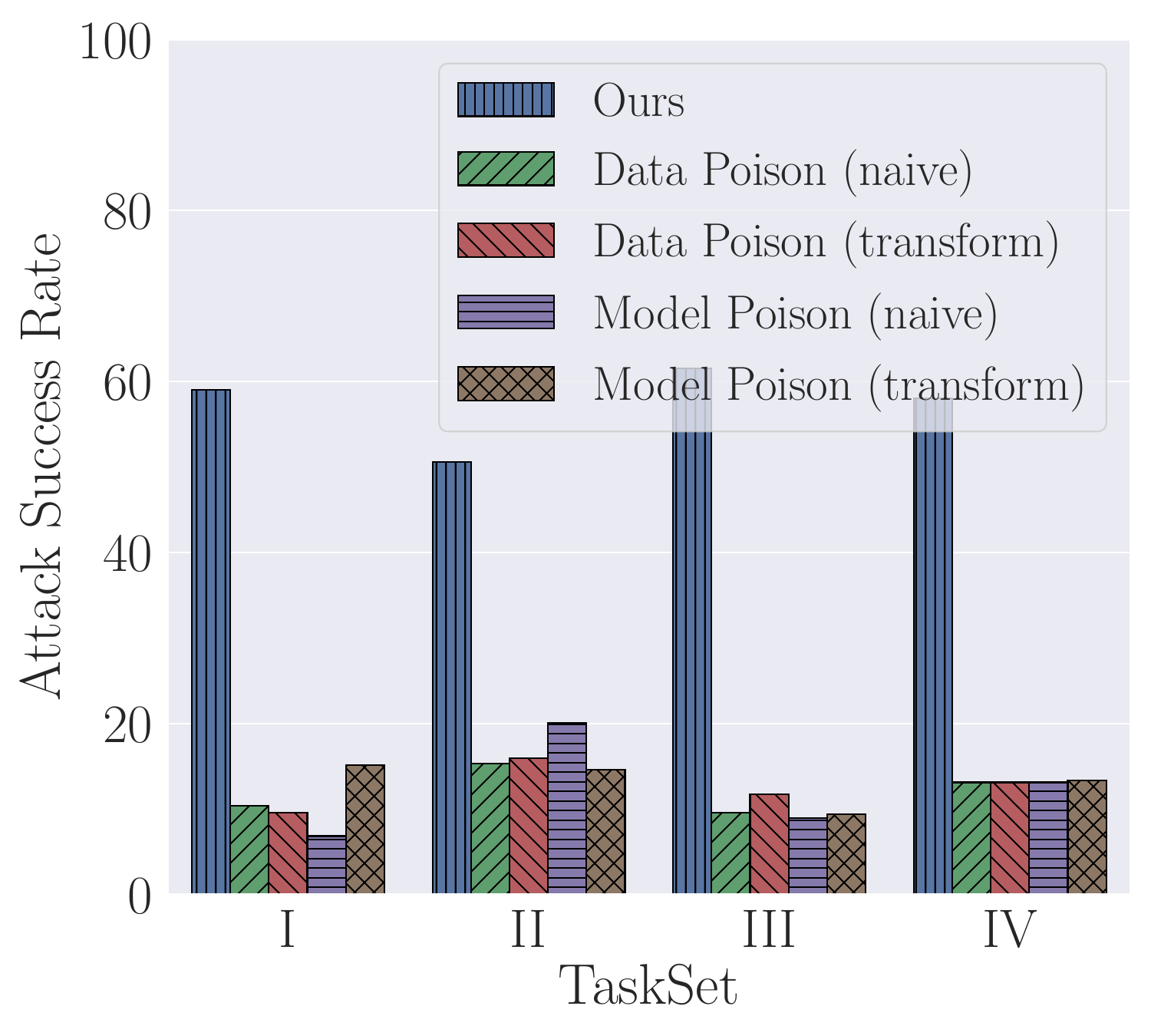}
\caption{MobileNet-V2}
\end{subfigure}
\begin{subfigure}{0.59\columnwidth}
\includegraphics[width=\columnwidth]{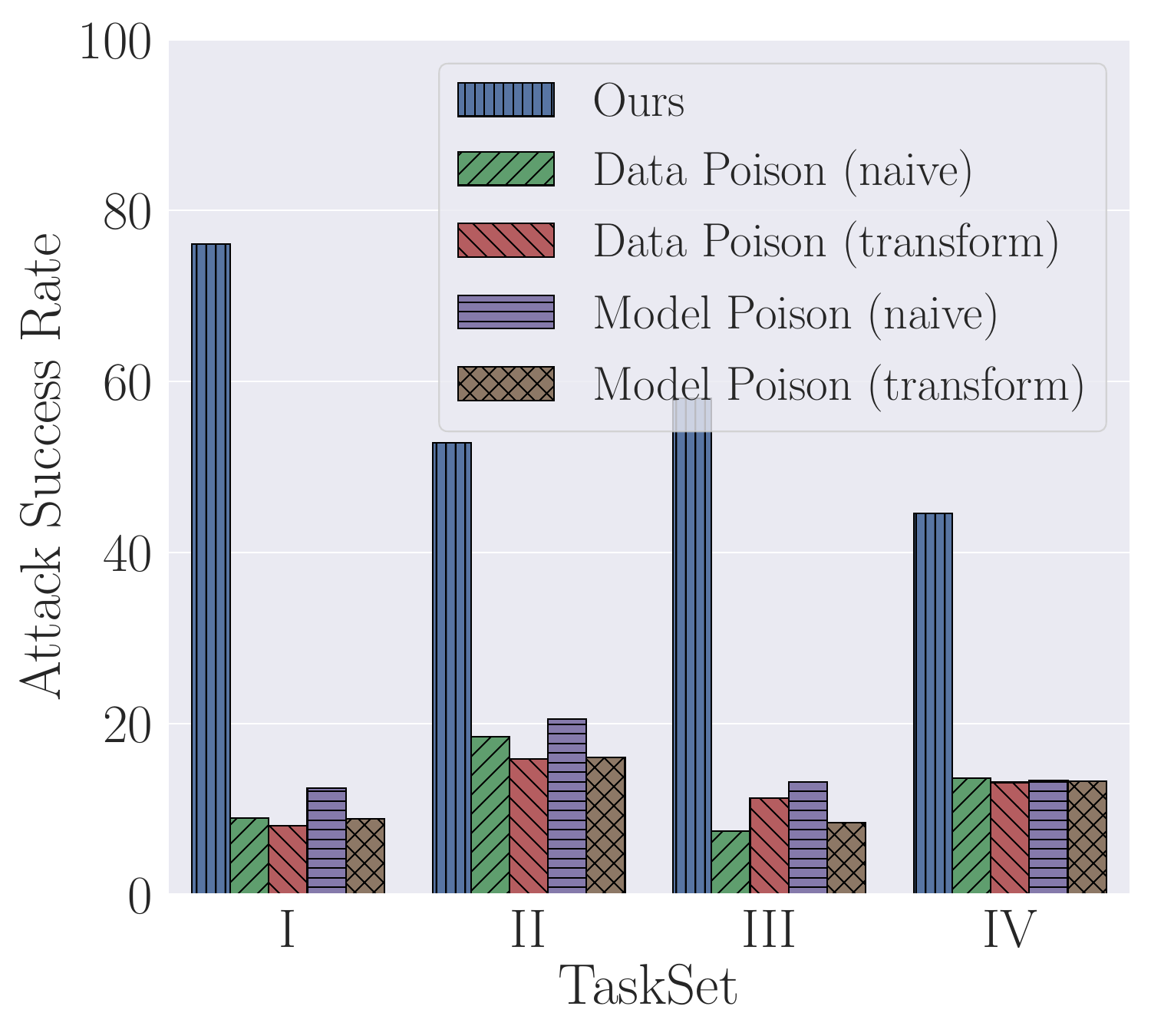}
\caption{ShuffleNet-V2}
\end{subfigure}

\caption{The attack success rates of \attack, the baseline attacks of data poison, and model poison.}
\label{fig:asr}
\end{figure*}

\mypara{Attack Success Rate} 
The attack success rate measures the hijacked global model performance on the hijacking testing dataset (\autoref{req2}).
We calculate the attack success rate by computing the accuracy of the hijacked global model on a hijacking testing dataset, which is from the same distribution as the hijacking training dataset.

%--------------------------------------------------
\section{Experimental Results}
%--------------------------------------------------
In this section, we conduct extensive experiments to evaluate the attack performance of \attack.
First, we present the utility and attack success rate of \attack, comparing them with baseline attacks. 
We then delve into the crucial roles of cloak computing and class mapping adopted in our approach. 
Finally, we explore the influence of various factors on the performance of \attack.

%--------------------------------------------------
\subsection{Attack Performance}\label{sec:attack_performance}
%--------------------------------------------------
\mypara{Utility}
We start by evaluating the utility of the hijacked global model on the original testing dataset. 
\autoref{table:utility} illustrates the utility of both the hijacked global models and the clean ones trained solely with original datasets. 
We observe that data poisoning (naive) and model poisoning (naive) induce slight side effects on the global model's utility to different degrees.
For instance, clean ResNet-18 achieves 88.03\% utility on TaskSet-\RNum{1}, while hijacked ResNet-18 achieves 86.39\% utility by model poison (naive). 
Conversely, data poison (transform), and particularly model poison (transform), exhibit a more significant negative impact on the global model's utility. 
For example, on TaskSet-\RNum{3}, clean MobileNet-V2 achieves 44.72\% utility, whereas hijacked MobileNet-V2 achieves 38.84\% and 29.94\% utility by data poison (transform) and model poison (transform), respectively. 
These findings align with those of Salem et al.~\cite{SBZ22}, where transforming the hijacking dataset leads to relatively poorer performance. 
We attribute these observations to the fact that transforming the hijacking dataset visually aligns it more closely with the original dataset, thereby complicating model optimization and resulting in decreased utility. 
In particular, since the model poison further amplifies poisoned local model parameters by the scale-up $\gamma=\frac n\eta $, it leads to a particularly obvious decrease in utility.

Encouragingly, \attack does not modify the local model, but only exchanges clean model parameters with the central server.
Thus, \attack does not affect the utility performance of the global model in the original task.
\attack can guarantee superior utility performance over baseline attacks.

\mypara{Attack Success Rate}
Now, we evaluate the attack success rate.
We also compare our attack with the baseline attacks of data poison and model poison.
We present the results in \autoref{fig:asr}.

As we can see, \attack achieves a very high level of attack success rate and is far superior to baseline attacks.
For example, on ResNet-18, our attack for TaskSet-\RNum{1} achieves an attack success rate of 92.75\%, while all baseline attacks achieve only around 10\%. 
The reason behind this is that the baseline attacks achieve their goal by submitting poisoned model parameters, thereby injecting the hijacking effect into the global model. However, as the global model undergoes more FL training rounds and averages only benign local model parameters, the hijacking effect will be reduced or even removed.
However, since \attack does not rely on injecting the hijacking effect into the global model, it has a high probability of achieving remarkable attack performance even as the global model undergoes more FL training rounds.

\begin{figure}[!t]
\centering
\begin{subfigure}{0.49\columnwidth}
\includegraphics[width=\columnwidth]{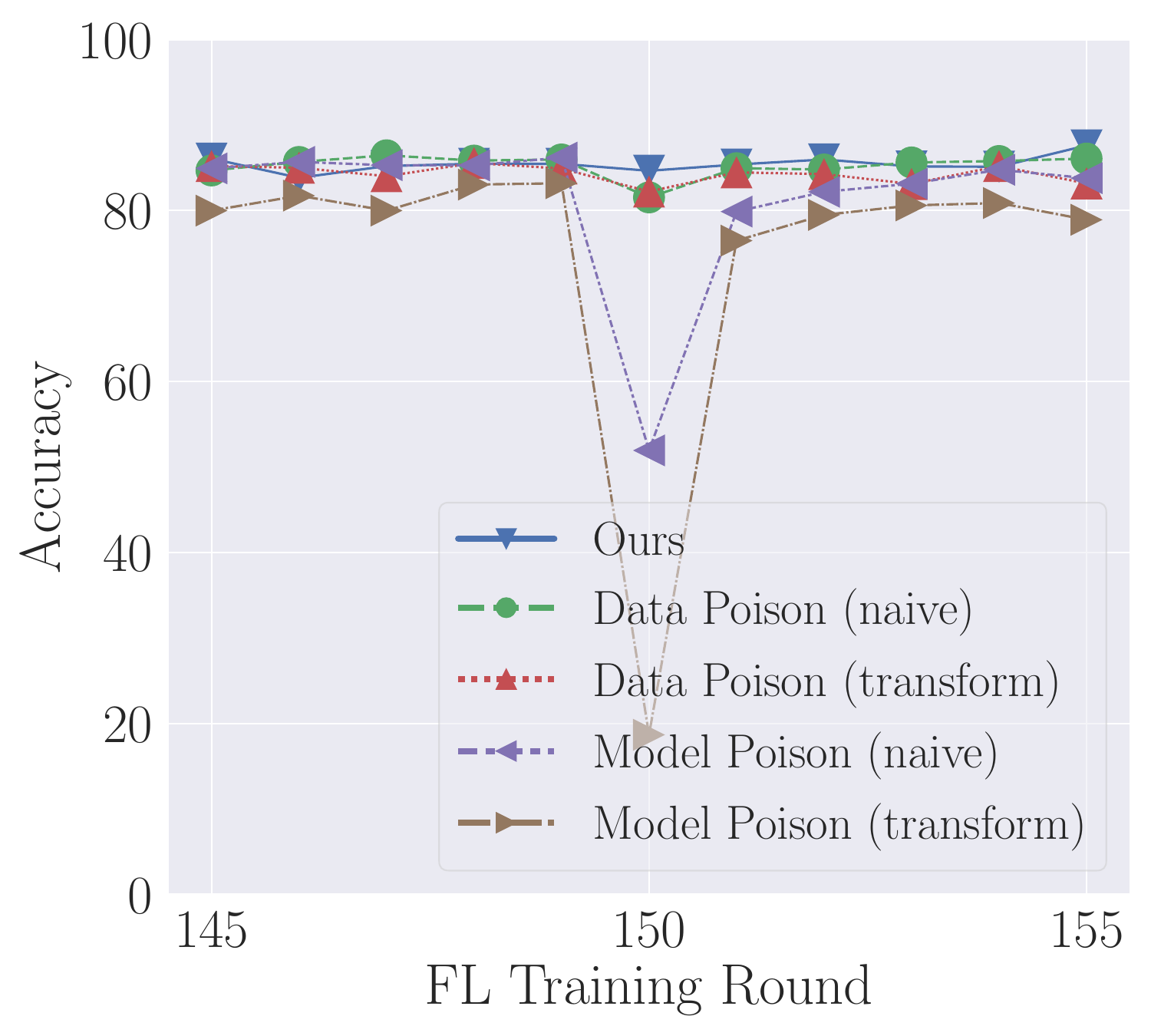}
\caption{ResNet-18}
\end{subfigure}
\begin{subfigure}{0.49\columnwidth}
\includegraphics[width=\columnwidth]{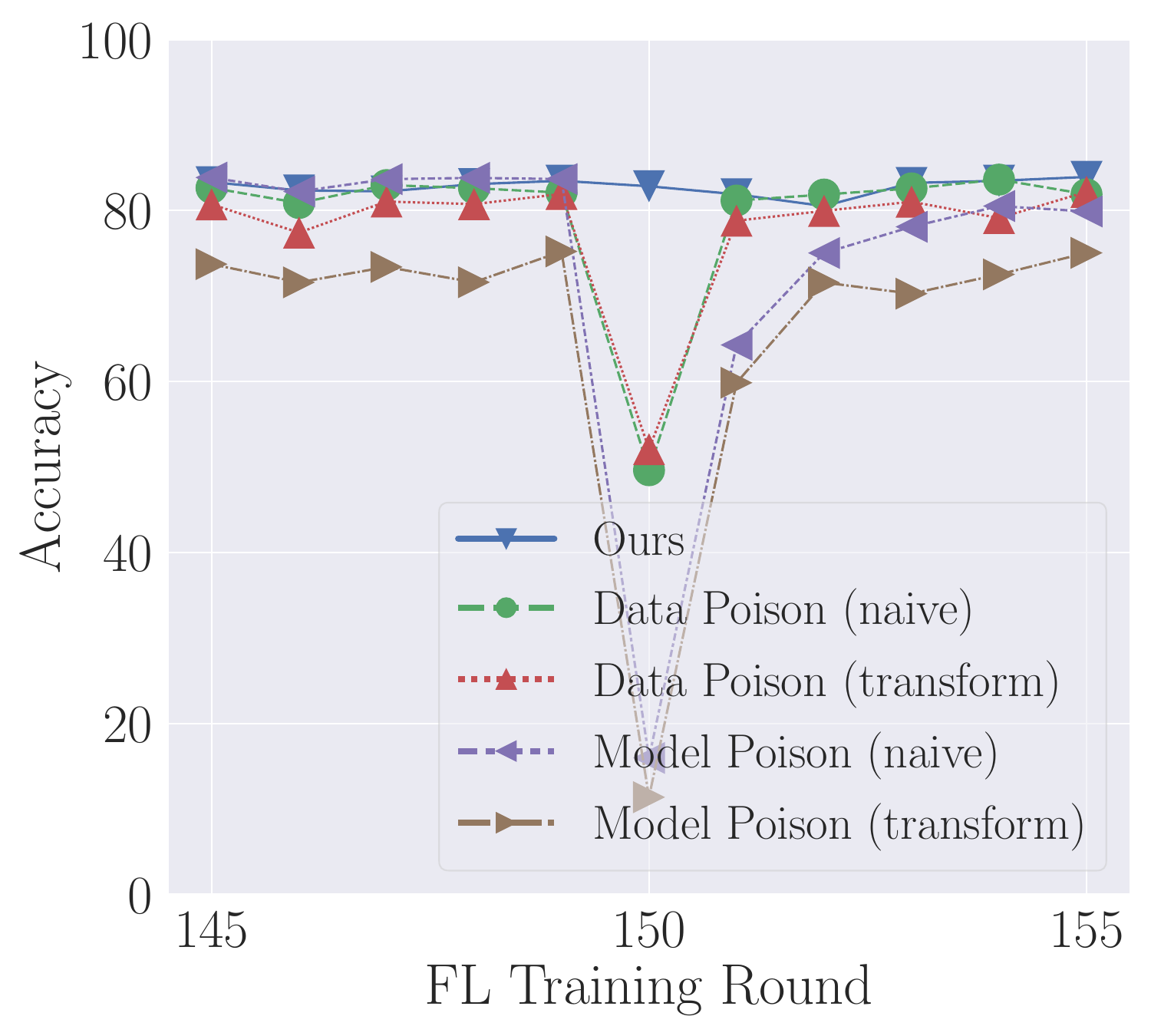}
\caption{MobileNet-V2}
\end{subfigure}
\caption{The utility of the global model before, during, and after the hijacking round (i.e., 150th).}
\label{fig:fluctuation}
\end{figure}

%--------------------------------------------------
\subsection{Fluctuation of Utility}
%--------------------------------------------------
Note that all evaluations, except those in this section, are performed on the final global model: the adversary starts cloak computation/data poison/model poison at the 150th training round and evaluates their attacks on the ultimate global model trained over the entire 200 rounds.

In this section, we investigate the impact of these attacks on the global model at the 150th hijacking round. 
\autoref{fig:fluctuation} illustrates the utility of the global model before, during, and after the hijacking round. 
We can observe a noticeable fluctuation in utility when baseline attacks execute the poison operation. 
Such fluctuations can easily alert the server, leading to detection by the central server. 
This observation stems from the fact that baseline attacks modify local model parameters, which are then transmitted to the central server. 
Consequently, the global model is affected, resulting in decreased accuracy on the original dataset. 
In subsequent rounds, when only benign local model parameters are submitted to the central server, the adversary's hijacking impact is eliminated. 
As a result, utility levels rebound to a high standard. 
This also explains why these baseline attacks only achieve a 5\% $\sim$ 22\% attack success rate when targeting the final global model, as depicted in \autoref{fig:asr}.
In contrast, in \attack, the adversary simply trains the local model on their original dataset as usual and submits these clean model parameters to the server. 
As a result, the adversary can always guarantee the utility of the global model.

%--------------------------------------------------
\subsection{Feature Visualization}
%--------------------------------------------------
It is important to note that the key idea of \attack is to add cloaks to the hijacking samples, which allows for the identification and extraction of features that are similar to the original samples.
In addition to the quantitative analysis described above, in this section, we will verify the effectiveness of \attack from a visualization point of view.

\begin{figure}[!t]
\centering
\begin{subfigure}{0.49\columnwidth}
\includegraphics[width=\columnwidth]{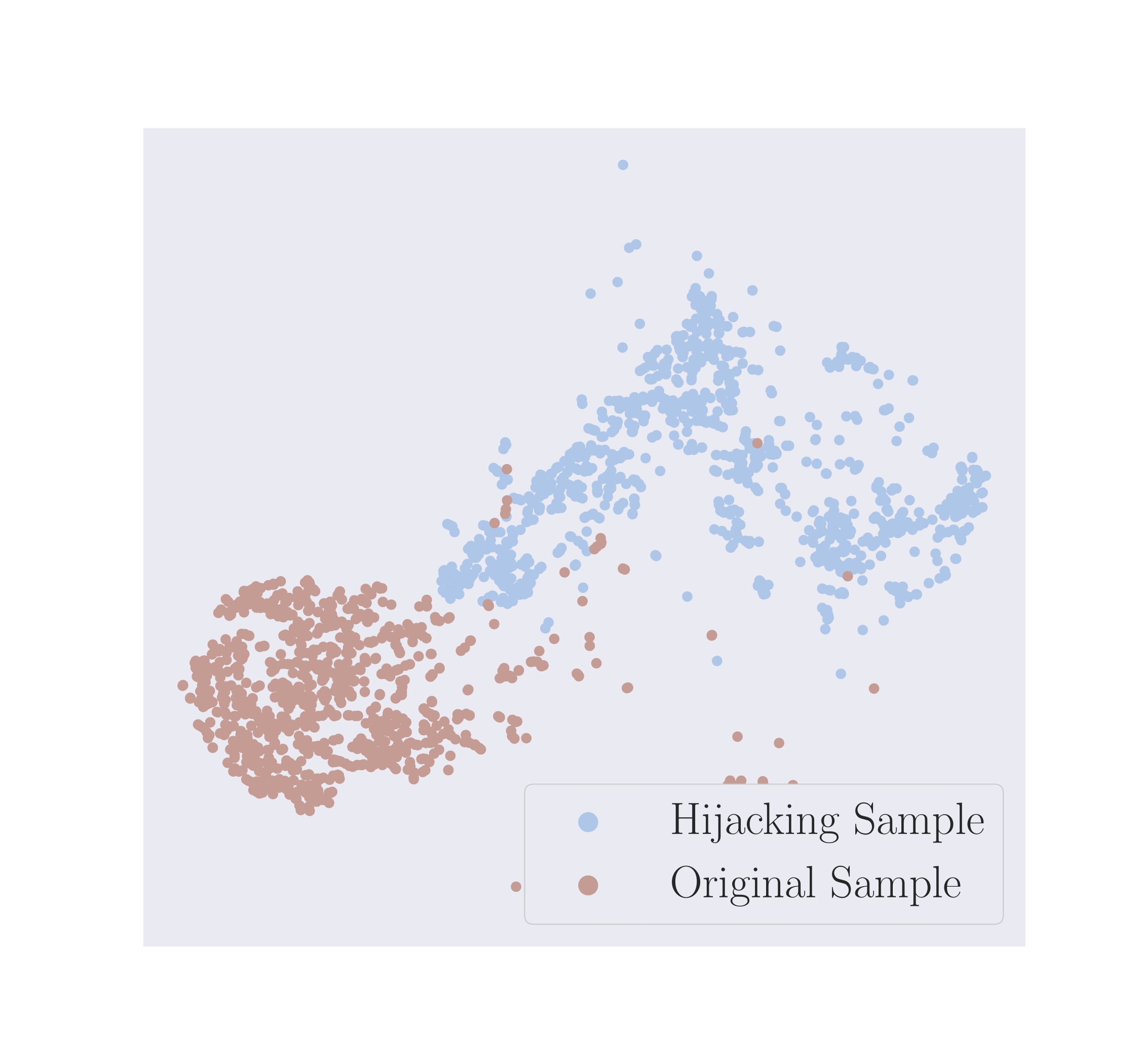}
\caption{Without Cloak}
\label{fig:vis_without_mask}
\end{subfigure}
\begin{subfigure}{0.49\columnwidth}
\includegraphics[width=\columnwidth]{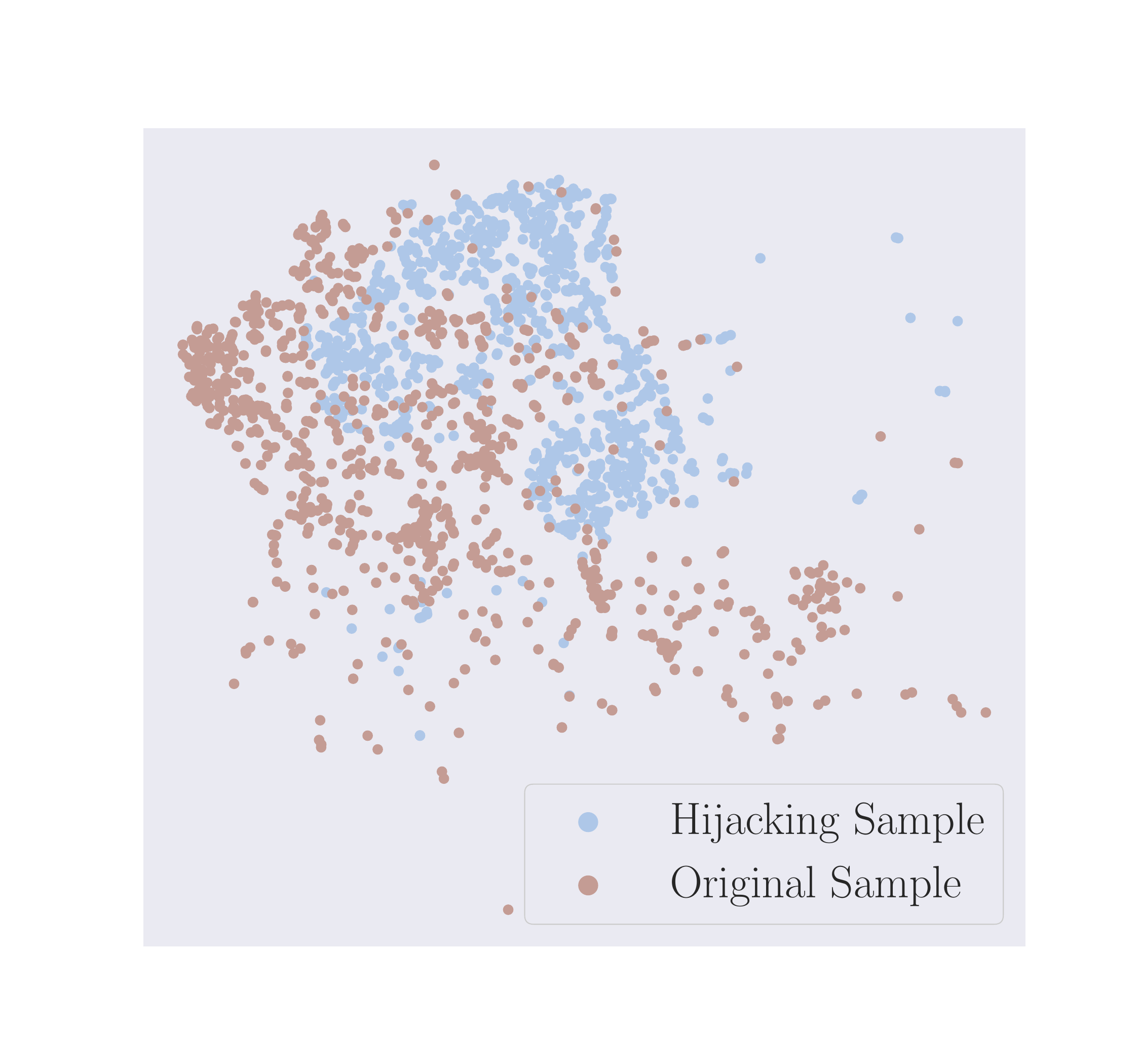}
\caption{With Cloak}
\label{fig:vis_with_mask}
\end{subfigure}
\caption{The visualization of hijacking/original sample features mapped to 2D space using t-SNE.}
\end{figure}

Concretely, in TaskSet-\RNum{1}, where CIFAR-10 is the original dataset, and MNIST is the hijacking dataset, we randomly select some hijacking samples from a single hijacking class and an equivalent number of original samples from its mapped original class. 
We feed these samples into the final global model and extract features from the middle layer of the model (i.e., $\Phi$ without the model classification layer).
Then, we utilize t-Distributed Stochastic Neighbor Embedding (t-SNE)~\cite{MH08} to process these features into a 2D space and visualize them in \autoref{fig:vis_without_mask}.
As we can see, the features from hijacking and original samples have been clearly mapped into separate regions.

In contrast, after we add the corresponding cloak to these hijacking samples, we then feed the cloaked samples and the original samples to extract their features and map these features in 2D space using t-SNE.
Remarkably, the visualization shown in \autoref{fig:vis_with_mask} demonstrates a high degree of overlap between the features of cloaked samples and the original samples. 
This intriguing finding suggests that the cloaked samples are now associated with features similar to the original samples, effectively leading the global model to classify these cloaked samples to this original class.

\begin{figure}[!t]
\centering
\begin{subfigure}{0.49\columnwidth}
\includegraphics[width=\columnwidth]{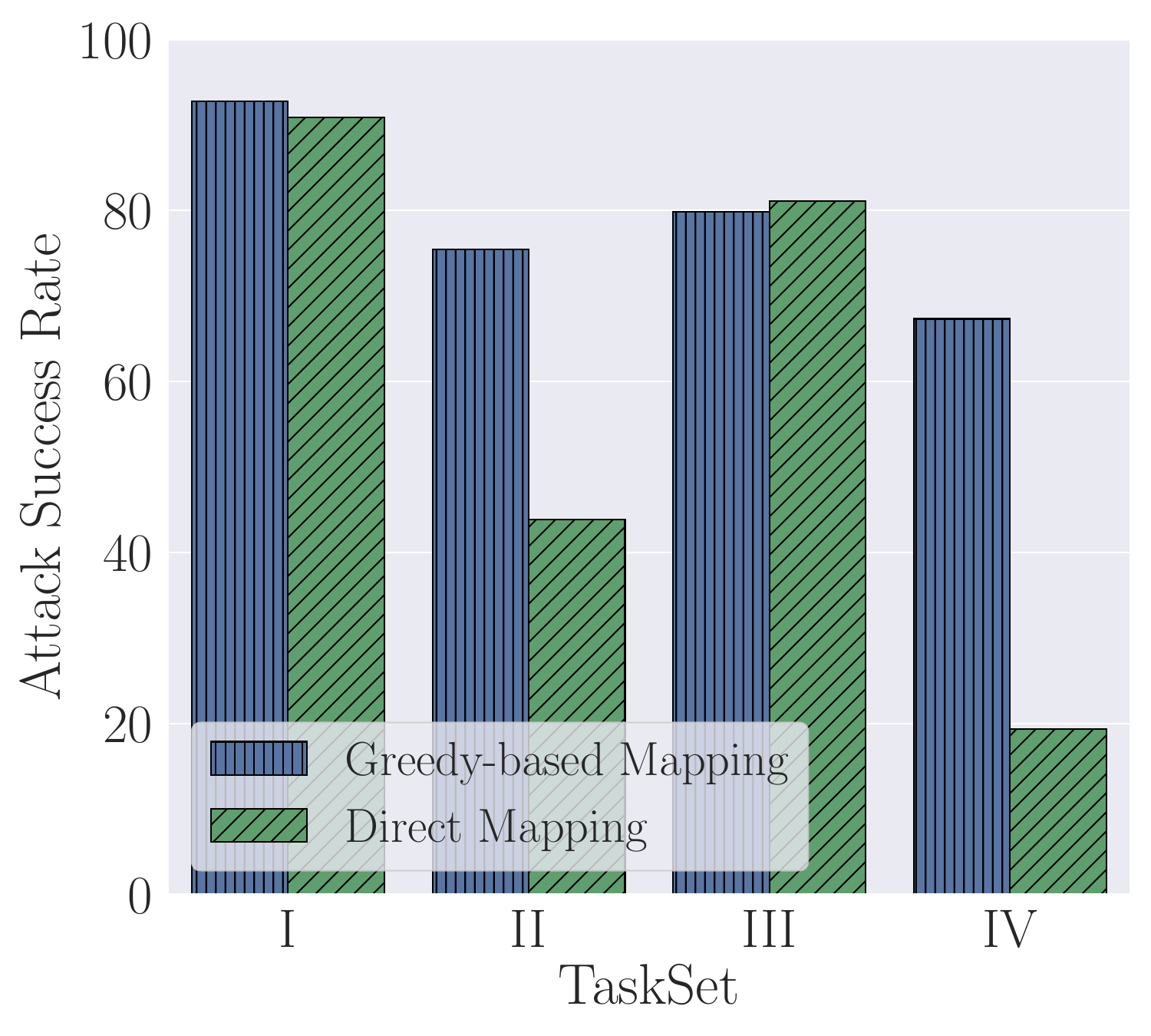}
\caption{ResNet-18}
\end{subfigure}
\begin{subfigure}{0.49\columnwidth}
\includegraphics[width=\columnwidth]{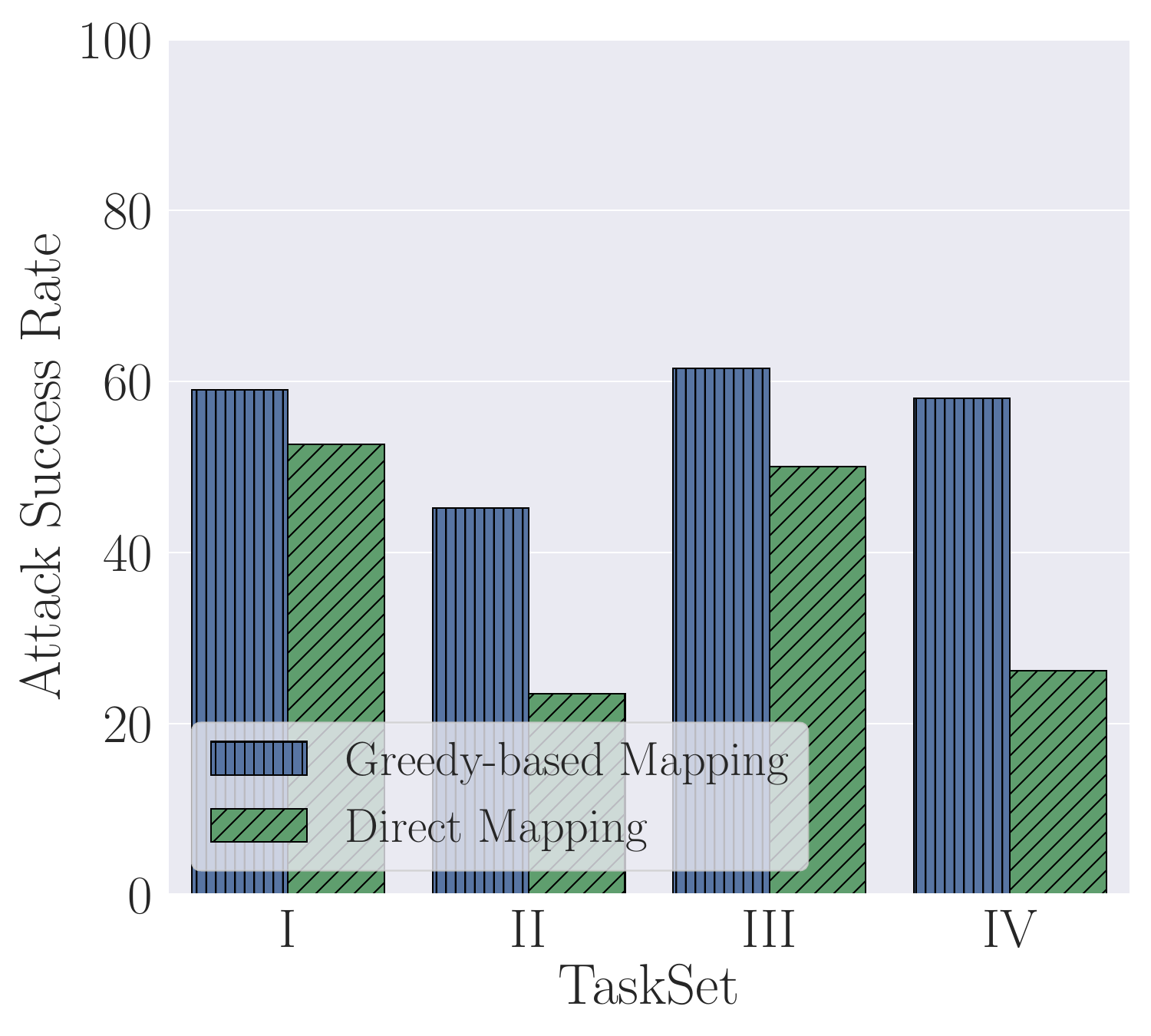}
\caption{MobileNet-V2}
\end{subfigure}
\caption{The attack success rate of \attack under the impact of the class mapping, i.e., greedy-based class mapping and direct class mapping.
}
\label{fig:class_mapping}
\end{figure}

%--------------------------------------------------
\subsection{Impact of Class Mapping}\label{sec:impact_class_mapping}
%--------------------------------------------------
Recall that our attack consists of four steps, where the first stage is class mapping. In this stage, the adversary predefines a class mapping between the hijacking class and the original class. 
Since we believe that the hijacking samples of certain classes will be more inclined to be associated with a certain original class. 
Therefore, we introduce a greedy-based class mapping to facilitate the optimization of effective cloaks. 
In this section, we investigate the impact of the class mapping on the attack performance, i.e., whether it is really as we conjectured.

In particular, we discard our proposed greedy-based class mapping strategy and instead assign the first several original classes to the hijacking classes in a one-to-one mapping, similar to the strategy adopted by Salem et al.~\cite{SBZ22}.
For example, in TaskSet-\RNum{3}, where TinyImageNet-100 serves as the original
dataset and MNIST as the hijacking dataset, so we directly assign the first 10 classes of TinyImageNet-100 to the 10 classes of MNIST in a one-to-one mapping (called direct mapping).
We then keep our (\attack) other settings unchanged and evaluate its attack performance.
\autoref{fig:class_mapping} shows a comparison of \attack with greedy-based class mapping and direct class mapping.
We can find that in most cases, greedy-based class mapping outperforms direct class mapping in terms of attack performance.
For example, in TaskSet-\RNum{2} on ResNet-18, the attack with greedy-based class mapping achieves an attack success rate of 75.45\%, while the attack with direct class mapping achieves an attack success rate of only 43.85\%.
However, we do find an exception where the direct mapping achieves a similar level of performance as greedy-based attacks in TaskSet-\RNum{3} on ResNet-18.
We attribute this to the fact that the hijacking samples of the first few classes are already more inclined to be associated with the first few original classes.
Thus, the direct class mapping also achieves a similar attack performance as the greedy-based class mapping.

In summary, the above results show that the greedy-based class mapping we designed has excellent performance, and also verify our claim that hijacking samples of certain classes will indeed be more inclined to be associated with a certain original class.

\begin{figure}[!t]
\centering
\begin{subfigure}{0.49\columnwidth}
\includegraphics[width=\columnwidth]{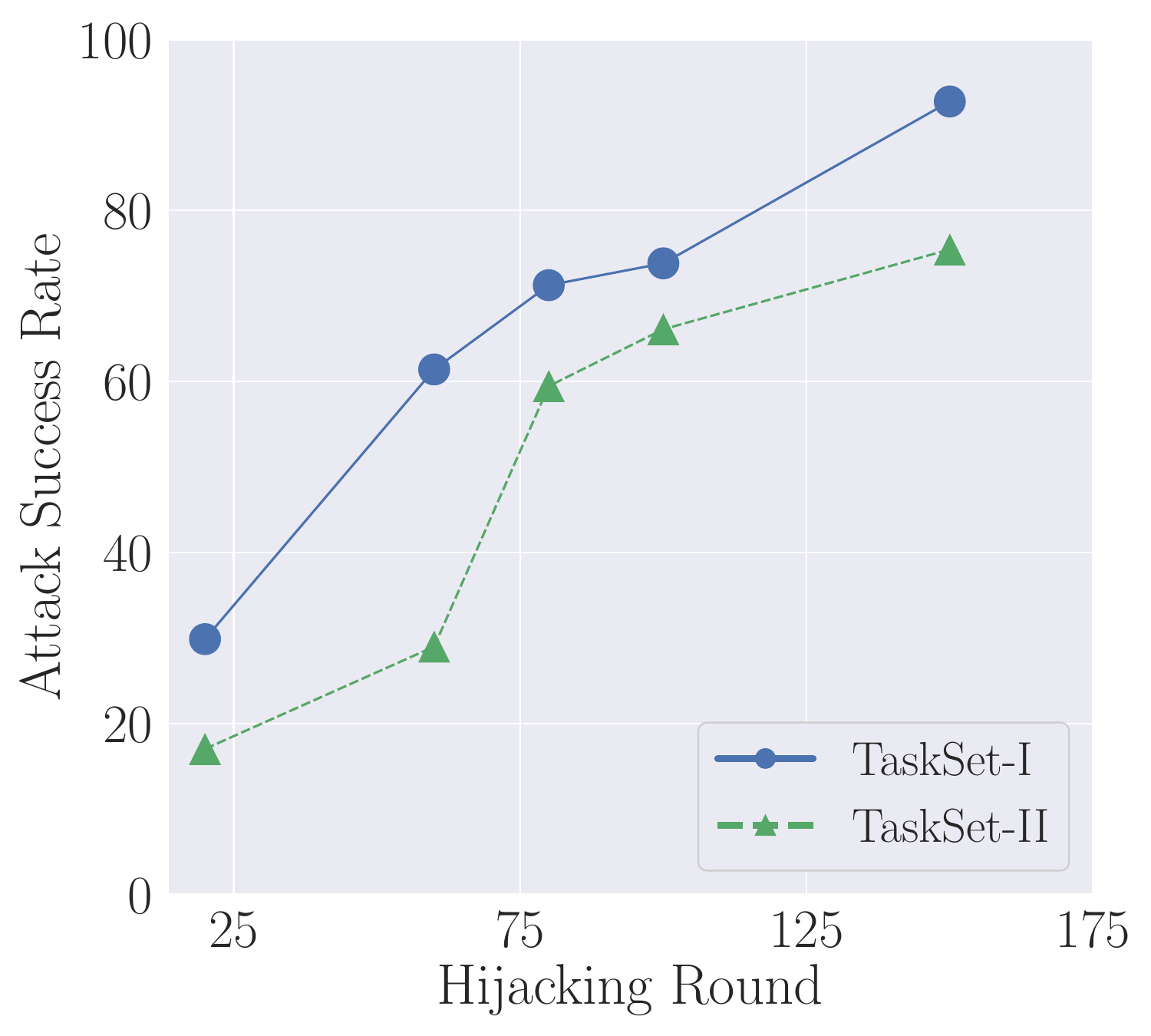}
\caption{ResNet-18}
\end{subfigure}
\begin{subfigure}{0.49\columnwidth}
\includegraphics[width=\columnwidth]{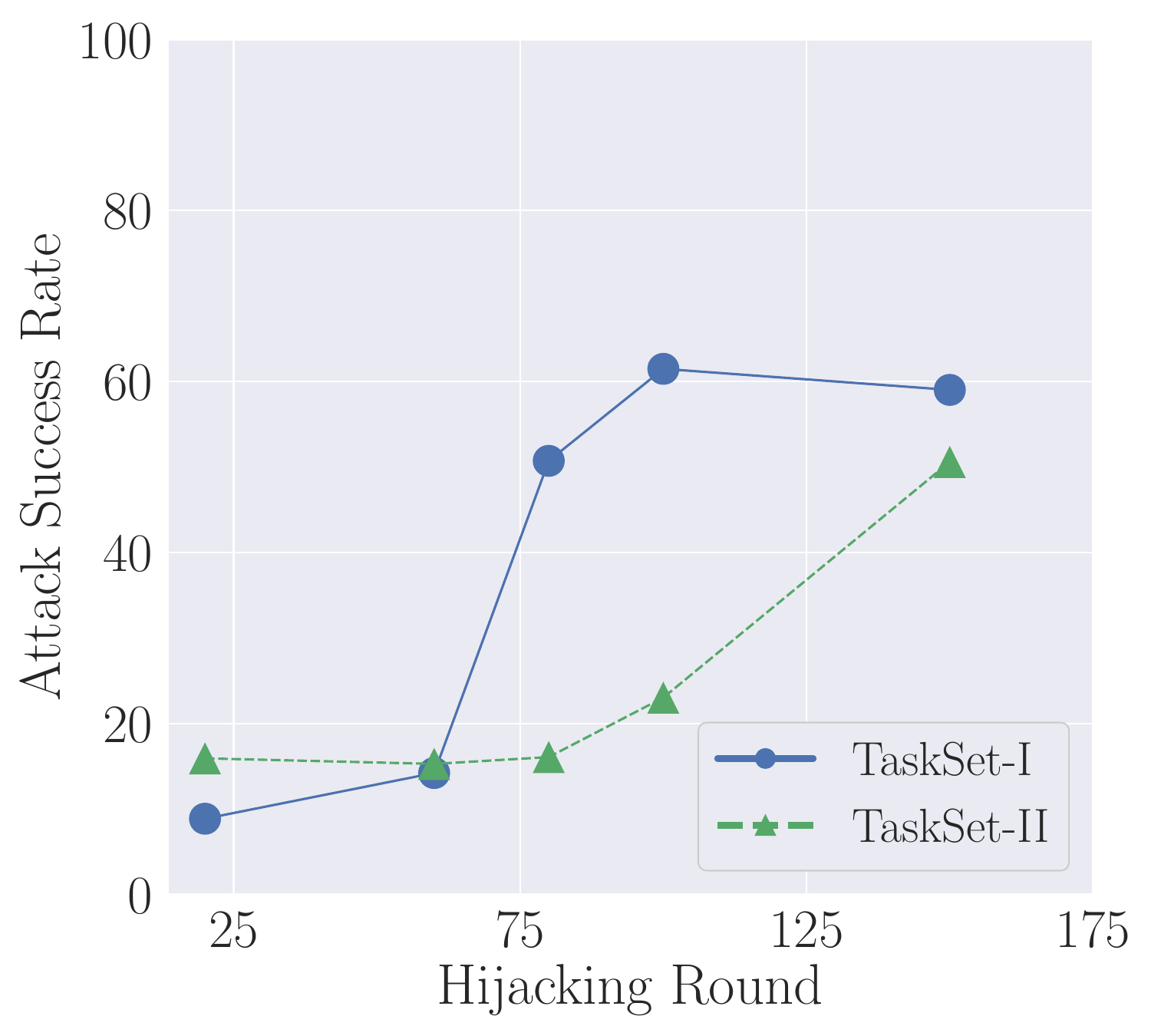}
\caption{MobileNet-V2}
\end{subfigure}
\caption{The attack success rate of \attack under the impact of the hijacking round.
}
\label{fig:hijacking_round}
\end{figure}

%--------------------------------------------------
\subsection{Ablation Study of Hyperparameters}
%--------------------------------------------------
\mypara{Hijacking Round} We now explore the impact of the hijacking round, denoting the FL training round where the adversary starts cloak computation. 
In the above evaluations, we default the adversary to start cloak computation at the $150$th training round. 
However, in real-world scenarios, the adversary may not have control over their participation round, as the central server might follow specific protocols, such as randomly selecting clients. 
Hence, it is essential to investigate how the hijacking round impacts our \attack.

Specifically, within the span of 200 FL training rounds, we configure the adversary to start cloak computation at the $10$th, $50$th, $80$th, $100$th, and $150$th training rounds, respectively. 
To clarify, for instance, the adversary starts cloak computation only once at the $80$th round and evaluates the attack on the ultimate global model trained over the entire 200 rounds.
Subsequently, we evaluate the efficacy of the obtained cloaks by launching attacks on the final global model.
As depicted in \autoref{fig:hijacking_round}, a clear trend emerges, indicating a positive correlation between the attack performance of \attack and the hijacking round. 
In other words, initiating cloak computation later in the adversarial process results in higher attack performance.

The underlying reason is that after the hijacking round, the global model undergoes further aggregation based on the submitted model updates in subsequent training rounds. This implies that the final global model differs from the one on which the adversary starts cloak computation. Such disparities impact the transferability of the cloaks. Starting cloak computation closer to the final training rounds, where the global model experiences minimal updates, allows the cloaked samples to be extracted to features similar to the original samples, resulting in a higher level of attack success rate.
In practical applications, the adversary can leverage its own original dataset to estimate the global model's convergence status. 
The adversary should initiate cloak computation when the global model has either converged or is close to convergence.

\begin{figure}[!t]
\centering
\includegraphics[width=0.8\columnwidth]{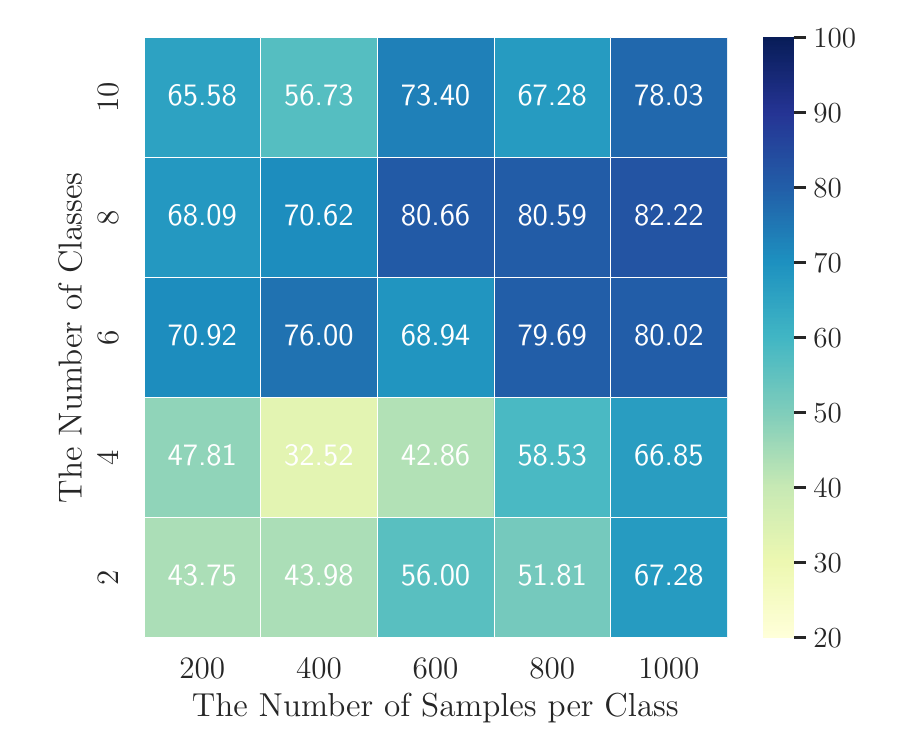}
\caption{The attack success rate under the impact of the complexity of the hijacking task. The y-axis represents the number of classes, and the x-axis represents the number of samples per class.
}
\label{fig:hijack_task_complexity}
\end{figure}

\mypara{Complexity of Hijacking Task}
We here study the impact of hijacking task complexity.
Specifically, we consider the hijacking task complexity from two perspectives: the number of classes and the number of samples per class. More concretely, for the original dataset TinyImageNet-100 and the hijacking dataset GTSRB, we reconstruct the GTSRB dataset by adjusting the number of classes from 2 to 10. Additionally, we vary the number of GTSRB samples per class within the range of 200 to 1000.
We conduct extensive experiments to simultaneously tune these two hyper-parameters and report the results in \autoref{fig:hijack_task_complexity}.
Through investigation, we make the following observations.
\begin{itemize}
    \item The attack performance tends to increase with the number of samples per class.
    \item The attack performance initially increases with the number of classes and then decreases, peaking around 6 or 8 classes.
\end{itemize}
The first observation is rooted in the fact that having more hijacking samples per class results in the learning of a more effective, general, and stable cloak, consequently leading to a higher attack performance.
Concerning the second observation, we speculate that it is linked to intricate interactions between the global model architectures and the complexity of the original task. 
We leave the in-depth exploration of such intricate interactions as a future work.

\mypara{Number of Hijacking Task}
We now discuss the impact of the number of hijacking tasks, which involves hijacking the global model with multiple datasets. Importantly, owing to our innovative design, \attack distinguishes itself by not interacting with the global model's parameters, in contrast to the baseline attacks of data poison and model poison. Consequently, the adversary can initiate cloak computation based on the same global model, irrespective of the number of hijacking datasets. This approach enables the adversary to possess distinct sets of cloaks for different hijacking datasets, offering flexibility when launching attacks on the final global model as they can select the suitable set of cloaks.

In contrast, it is expected that the baseline attacks of data poison and model poison will be influenced by the number of hijacking tasks. More hijacking datasets are likely to result in submitted model updates deviating further from a clean one, thereby causing a larger utility drop in the global model.

\begin{figure}[!t]
\centering
\begin{subfigure}{0.49\columnwidth}
\includegraphics[width=\columnwidth]{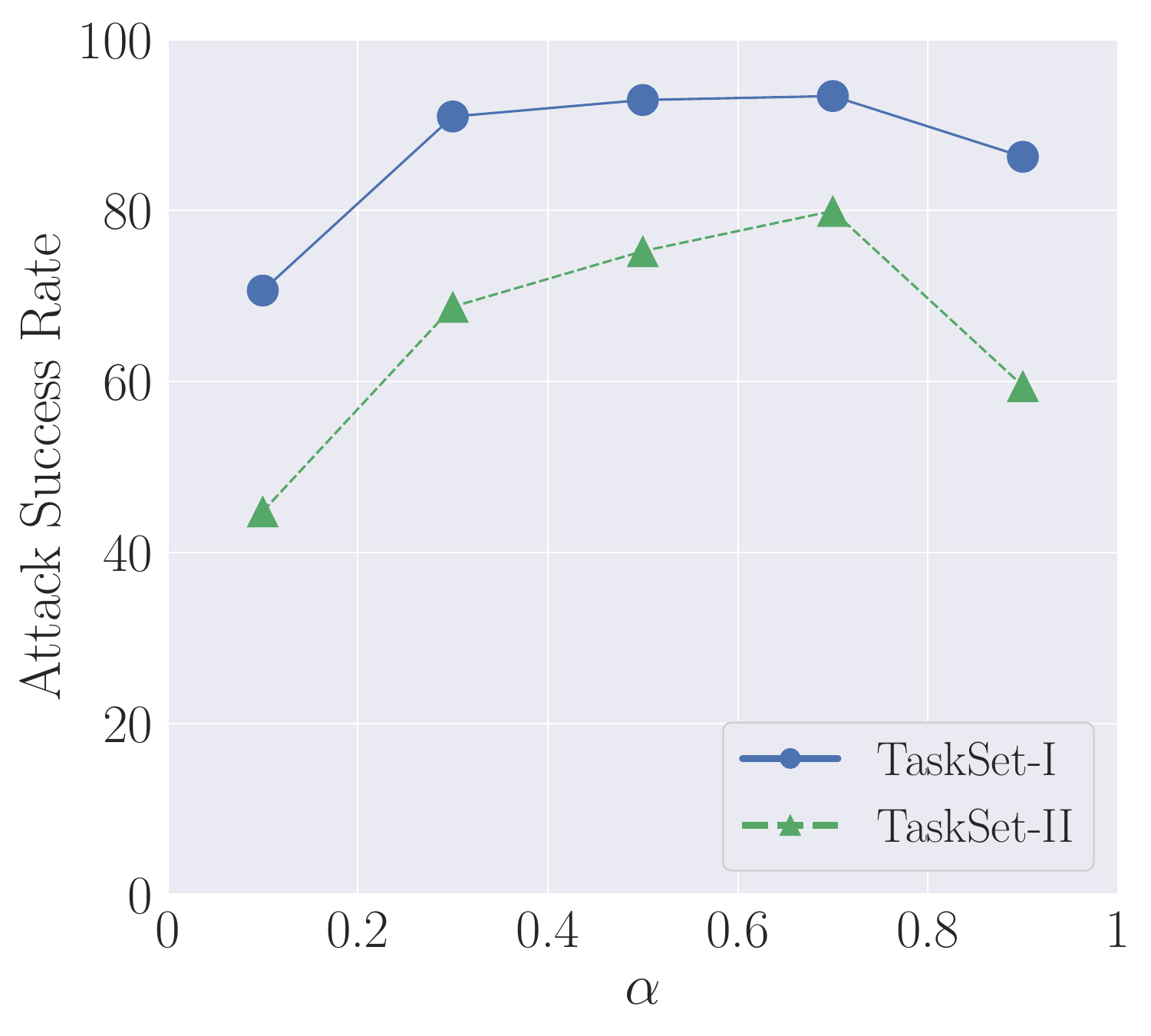}
\caption{ResNet-18}
\end{subfigure}
\begin{subfigure}{0.49\columnwidth}
\includegraphics[width=\columnwidth]{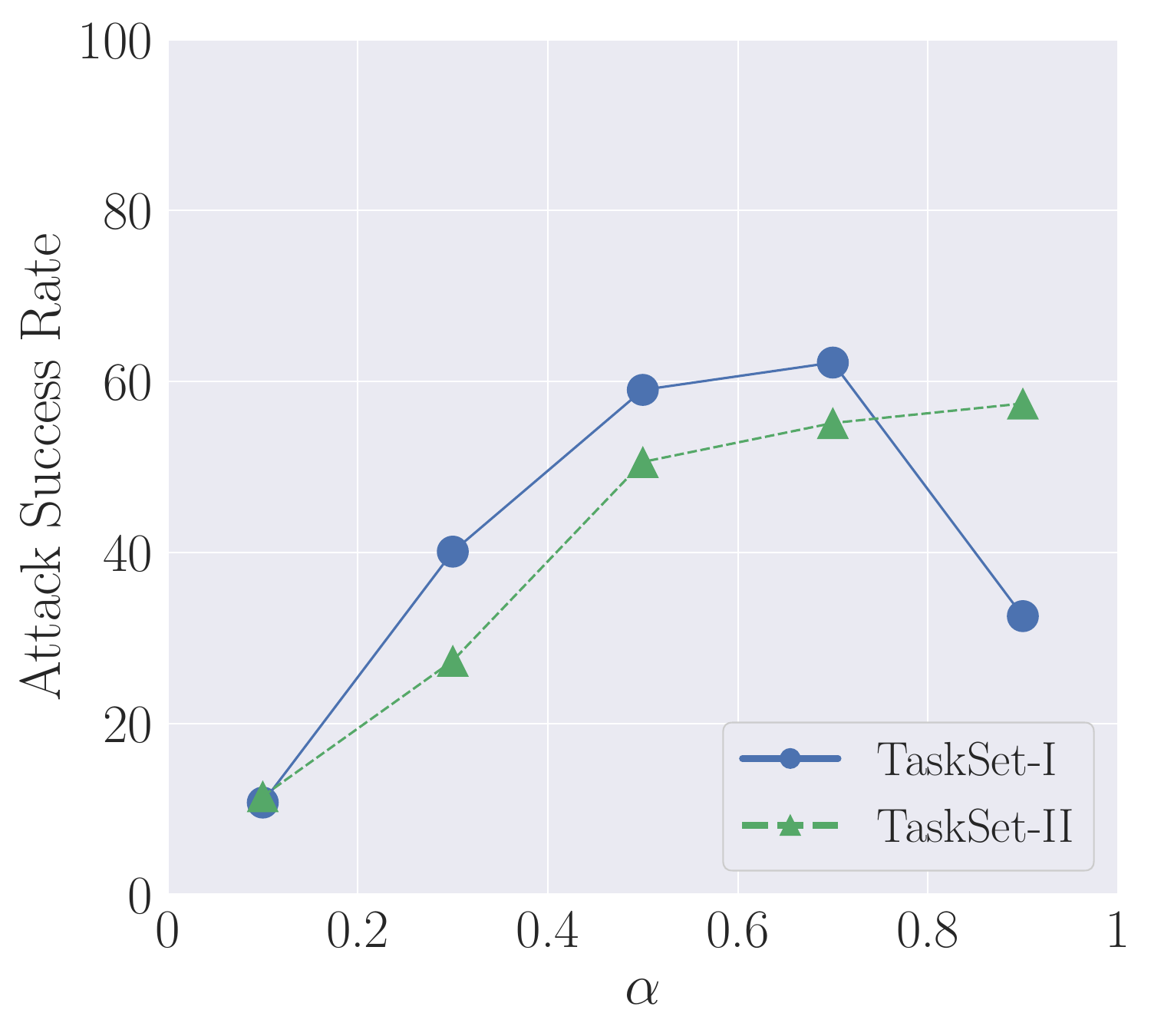}
\caption{MobileNet-V2}
\end{subfigure}

\caption{The attack success rate of \attack under the impact of $\alpha$}
\label{fig:alpha}
\end{figure}

\mypara{Balance Parameter $\alpha$}\label{sec:alpha}
Recall that we employ a convex combination of the hijacking samples and the cloak: ${x_h}\oplus \delta_h = \alpha{x_h} +(1-\alpha ) \delta_h $. In this expression, the parameter $\alpha \in [0, 1]$ governs the balance between the weight or influence of $x_h$ and $\delta_h$ on the cloaked samples. In this section, we investigate the impact of $\alpha$ on the attack performance. Note that we use a default value of 0.5 for $\alpha$ in the evaluations presented above.

Specifically, we vary the parameter $\alpha$ from 0 to 1 and present the results in \autoref{fig:alpha}. We clearly observe a trend where the attack performance initially increases and then decreases in most cases. The peak attack performance is attained when $\alpha$ falls within the range of 0.5 to 0.7.
This observation motivates the adversary to either carefully select the optimal $\alpha$ or simply use a value within the range of 0.5 to 0.7.

\mypara{Number of Cloak} In our previous evaluations, we employ a single generalized cloak for all hijacking samples within the same class. 
In this section, we explore the impact of the number of cloaks on attack performance. 
In particular, we generate only one cloak for all hijacking samples, irrespective of their respective hijacking classes. 
\autoref{tab:num_cloak} shows evaluation results on ResNet-18 and MobileNet-V2 with all TaskSets.
We can clearly observe that employing just one cloak for the entire hijacking classes fails to effectively achieve the hijacking goal. 
This outcome is understandable, as it becomes considerably more challenging for a single cloak to establish class mappings from various hijacking classes to various original classes compared to the simpler task of mapping one hijacking class to one original class.

We further consider the feasibility of generating a specific cloak for each hijacking sample. 
We assert that this approach is impractical for several reasons. 
Firstly, optimizing a cloak for each individual sample limits its effectiveness to that sample alone, rendering it ineffective for new, unseen hijacking instances. 
Moreover, in hijacking datasets with numerous samples, generating a unique cloak for each becomes unwieldy for the adversary, making selection for new samples challenging.
Thus, adversaries should optimize cloaks based on a data domain (e.g., hijacking samples within the same class) with the aim of achieving generalizability.

\begin{table}[!t]
    \centering
\caption{Comparison between \attack that uses only one cloak for the entire hijacking dataset and \attack that employs one cloak for each hijacking class (i.e., multi cloaks).}\label{tab:num_cloak}
\scalebox{0.7}
{
\begin{tabular}{l|c|c|c|c}
\toprule
\multirow{2}{*}{TaskSet} & \multicolumn{2}{c|}{ResNet-18} & \multicolumn{2}{c}{MobileNet-V2} \\
\cmidrule(lr){2-5}
                       & One Cloak      & Multi Cloak     & One Cloak      & Multi Cloak       \\
\cmidrule(lr){1-5}
\RNum{1}               & 0.1246         & 0.9275      &      0.1246       & 0.5903        \\
\cmidrule(lr){1-5}
\RNum{2}               &     0.1200     & 0.7545      &         0.1200     & 0.5060        \\
\cmidrule(lr){1-5}
\RNum{3}               &      0.1135      & 0.7984      &    0.1135        & 0.6156        \\
\cmidrule(lr){1-5}
\RNum{4}               &       0.1111     & 0.6733      &       0.1111       & 0.5808     \\
\bottomrule  
\end{tabular}
}
\end{table}

%--------------------------------------------------
\section{Discussion}
%--------------------------------------------------
In this section, we discuss the possible defenses and the limitations of our attack.

%--------------------------------------------------
\subsection{Possible Defenses}
%--------------------------------------------------
Since detecting attacks during the FL training phase is impossible, we explore two defenses in the FL prediction phase: feature-based anomaly detection and adversarial example detection.

\begin{figure}[!t]
\centering
\includegraphics[width=0.8\columnwidth]{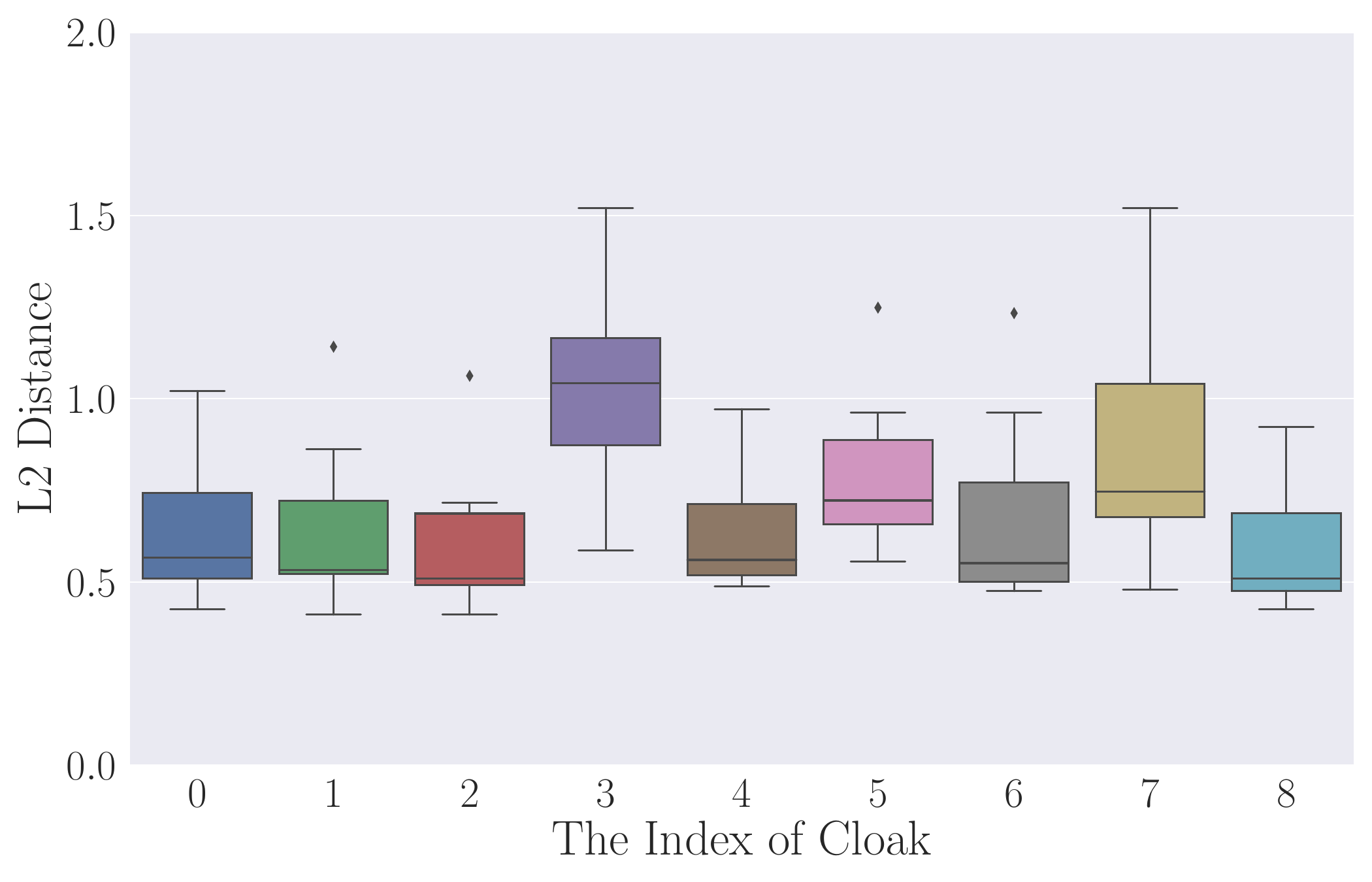}
\caption{The L2 distance between the cloaked sample's feature and the negative anchor feature. The x-axis represents the index of different cloaks added to the hijacking sample.}
\label{fig:defense}
\end{figure}

\mypara{Feature-based Anomaly Detection}
The intuition is motivated by \attack's execution stage: given a hijacking sample for query, the adversary will add all cloaks for it and submit these cloaked samples to the targeted deployed global model. 
Among these queries, only one cloak will map the given sample to the original class $y$, while all other cloaks will consistently map the given sample to the negative original class $y^{\ast}$.
As a result, there will be one cloaked sample with feature significantly further away from the negative anchor feature, compared to the other cloaked samples, as depicted in \autoref{fig:defense}.

\begin{table}[!t]
    \centering
\caption{The attack success rate of \attack against the feature-based anomaly defense.}\label{tab:feature_based_defense}
\scalebox{0.7}
{
\begin{tabular}{l|c|c|c|c}
\toprule
\multirow{2}{*}{TaskSet} & \multicolumn{2}{c|}{ResNet-18} & \multicolumn{2}{c}{MobileNet-V2} \\
\cmidrule(lr){2-5}
                         & No Defense      & Defense     & No Defense       & Defense       \\
\cmidrule(lr){1-5}
\RNum{1}            & 0.9275                & 0.1164       &            0.5903           & 0.1150        \\
\cmidrule(lr){1-5}
\RNum{2}                         &        0.7545     & 0.1158     &         0.5060     & 0.1108       \\
\cmidrule(lr){1-5}
\RNum{3}                        &      0.7984      & 0.1073     &    0.6156        & 0.1009       \\
\cmidrule(lr){1-5}
\RNum{4}                         &       0.6733     & 0.1041      &       0.5808       & 0.1088     \\
\bottomrule  
\end{tabular}
}
\end{table}

Building upon the above insight, we propose a feature-based anomaly detection.
Precisely, the defense pipeline consists of the following steps:
\begin{enumerate}
    \item The model owner first generates a set of anchor features regarding each original class (\autoref{anchorfeature}).
    \item When an adversary queries a set of cloaked samples (i.e., a hijacking sample with different cloaks), the model owner extracts their features.
    \item For each anchor feature, the model owner measures the L2 distance between it and each cloaked sample's feature.
    \item If only one cloaked sample's feature is much far away from the anchor feature (i.e., L2 distance larger than a threshold, see Appendix \autoref{tab:threshold_feature_defense}), the queried set of samples is considered to serve a hijacking sample.
\end{enumerate}
\autoref{tab:feature_based_defense} reports the attack performance under the feature-based anomaly detection.
Note that this defense is specifically designed for a set of cloaked samples serving hijacking samples and has no side effects on the utility, so we do not report it.
We can see that this defense can substantially mitigate the attack.
However, it's important to note that the adversary may not query a set of cloaked samples all at once; instead, they can intelligently query them at different times.
Furthermore, in the real world, the model is typically queried with mostly original samples, and interspersed among them are some cloaked samples.
Both of the above situations present significant challenges for this defense.
We leave the in-depth exploration of more effective defense mechanisms against our attack as a future work.

\mypara{Adversarial Example Defenses}
\attack adds cloaks to the hijacking samples, similar to adversarial examples that add slight pixel-level perturbations to input samples. 
Therefore, a straightforward defense strategy is utilizing adversarial example defense. 
Here, we employ a widely used adversarial example defense called Feature Squeezing~\cite{XEQ18}, which detects adversarial examples at the model prediction phase.
This defense is driven by the observation that the input spaces are often unnecessarily large, and this vast input space provides extensive opportunities for an adversary to construct adversarial examples. 
Thus, the defender can ``squeeze'' out unnecessary input space to reduce the degrees of freedom available to an adversary.
The key idea is to compare the model’s prediction on the input sample with its prediction on the sample after squeezing.
If the input sample and squeezed samples produce substantially different outputs from the model, the input sample is likely to be adversarial. 
By comparing the difference between predictions with a selected threshold, the defender can reject adversarial inputs.

Since this defense also squeezes the original sample, we report the utility of the global model in addition to the attack performance, as shown in \autoref{tab:defense_FS}.
In particular, we examine two thresholds (see Appendix \autoref{tab:thres_fs_defense}), set significantly lower/higher, to increase sensitivity/insensitivity to input examples with added perturbations, respectively. 
We find that a smaller threshold (Threshold 1) successfully detects hijacking samples but also yields a high rate of false positives when detecting original samples, resulting in a significant degradation of utility. 
Conversely, with a higher threshold (Threshold 2), this defense fails to detect hijacking, leading to \attack still outperforming baseline attacks.

\begin{table}[!t]
    \centering
\caption{The defensive performance of adversarial example defense, i.e., Feature Squeezing.}\label{tab:defense_FS}
\scalebox{0.65}
{
\begin{tabular}{l|c|c|c|c|c|c|c|c}
\toprule
\multirow{4}{*}{TaskSet} & \multicolumn{4}{c|}{ResNet-18}                                      & \multicolumn{4}{c}{MobileNet-V2}                                   \\
\cmidrule(lr){2-9}
                         & \multicolumn{2}{c|}{Threshold 1} & \multicolumn{2}{c|}{Threshold 2} & \multicolumn{2}{c|}{Threshold 1} & \multicolumn{2}{c}{Threshold 2} \\
\cmidrule(lr){2-9}
                         & Utility           & ASR          & Utility          & ASR          & Utility           & ASR          & Utility           & ASR         \\
\cmidrule(lr){1-9}
\RNum{1}                       &       0.5736            &       0.1223       &          0.8641        &      0.8934        &          0.6126         &        0.1397      &       0.8423      &    0.5304      \\
\cmidrule(lr){1-9}
\RNum{2}                        &      0.5776             &        0.1267      &           0.8633       &      0.7128        &        0.6090           &      0.1125     &     0.8428      &    0.4260      \\
\cmidrule(lr){1-9}
\RNum{3}                        &      0.2390             &       0.1179       &         0.4966         &      0.5771        &         0.3140          &       0.1081       &       0.4482        &     0.5417    \\
\cmidrule(lr){1-9}
\RNum{4}                        &        0.2402           &       0.1206       &        0.4966          &       0.5543       &          0.3108         &         0.0894     &     0.4484        &   0.4452  \\       
\bottomrule  
\end{tabular}
}
\end{table}

%--------------------------------------------------
\subsection{Limitation}\label{sec:limitation}
%--------------------------------------------------
The limitation of our \attack lies in the constraint that the hijacking dataset cannot contain more classes than the original dataset. This restriction is prevalent across all existing studies on model hijacking attacks. In our evaluation, all hijacking datasets had fewer classes compared to their corresponding original datasets. For instance, when considering CIFAR-10 as the original dataset, the hijacking dataset derived from MNIST or GTSRB could only encompass a maximum of 9 classes. 
The reason for this is that we need to reserve one original class to apply to all negative pairs, while there are only 9 original classes that can be assigned to the hijacking dataset.
We plan to explore more effective relabeling to address this limitation in future work.

The second limitation is that the cloaked samples queried by the adversary to the global model may exhibit noticeable visual differences from the original samples, making them susceptible to detection by the defense mechanisms. One potential solution is to not only minimize the deviation of features between the cloaked samples and original samples but also minimize the deviation of pixels in the image space. This approach aims to make the hijacking samples, added with cloaks, resemble original samples in both image space and feature space. However, we speculate that introducing this additional objective for cloaks may diminish its overall effectiveness. 
We will explore a more powerful attack against FL models to address this limitation.

%--------------------------------------------------
\section{Related Work}
%--------------------------------------------------

%--------------------------------------------------
\subsection{Adversarial Reprogramming}
%--------------------------------------------------
Adversarial reprogramming~\cite{EGS19,englert2022adversarial} is a powerful attack technique that aims to reprogram ImageNet classifiers, enabling them to perform tasks differently from their original design. 
Unlike the model hijacking attack, which operates during the training phase, adversarial reprogramming is a test-time attack. 
In this attack, the adversary optimizes a program that serves as the input to the target model, effectively tricking the model into performing a different task.

\attack and adversarial reprogramming are similar, involving the crafting of inputs by introducing perturbations to deceive the target model. However, adversarial reprogramming crafts perturbations based on the white-box access to the target model being attacked, which is unrealistic in the real world. In contrast, \attack initiates cloak computation based on an intermediate version of the global model and launches the attack against the final global model. Besides, the perturbation in \attack is class-dependent, meaning that hijacking samples of different classes use different perturbations. In contrast, adversarial reprogramming uses a universal perturbation for all samples of the hidden task. We also conducted tests on this method, and it demonstrated unacceptable attack performance (see ``Number of Cloak'').

%--------------------------------------------------
\subsection{Data Poisoning Attack}
%--------------------------------------------------
The data poisoning attack~\cite{BNL12,JOBLNL18,STLLXCS18} is a training time attack where the adversary manipulates the training process by inserting malicious data into the training dataset of the target model, with the intention of compromising the model's utility. 
The adversary typically achieves data poisoning by flipping the ground truth labels or injecting poisoned training datasets, causing the target model to learn incorrect patterns.
This attack has been studied in various machine learning domains, including federated learning, support vector machines (SVM), and regression learning. 

%--------------------------------------------------
\subsection{Backdoor Attack}
%--------------------------------------------------
The backdoor attack is typically associated with a specific trigger, such that when the trigger is present in any input sample, the target model predicts a predetermined label. 
The first backdoor attack introduced in machine learning was BadNets by Gu et al.~\cite{GDG17}, where a white square placed in the corner of an image served as the trigger to misclassify backdoored inputs. 
Subsequent variations of the backdoor attack, such as dynamic backdoors~\cite{SWBMZ22} and Trojan attacks~\cite{LMALZWZ18}, have been proposed to enhance the stealthiness and effectiveness of the attack.
Bagdasaryan et al.~\cite{BVHES20} have demonstrated that federated learning models are also susceptible to backdoor attacks.

%--------------------------------------------------
\subsection{Model Hijacking Attack}
%--------------------------------------------------
Model hijacking attacks have emerged as a novel training time attack technique that aims to repurpose the target model to perform a specific task defined by the adversary. The attack was first demonstrated by Salem et al.~\cite{SBZ22}, who successfully hijacked image classifiers to perform alternative image classification tasks. 
For example, they hijacked models trained on datasets like CIFAR-10 or CelebA using the MNIST dataset as the hijacking dataset.
Recently, Si et al.~\cite{LLHYBZ222} broadened the scope of this attack to include text generation and classification models, hence showing its broader applicability.
However, all existing model hijacking attacks are only designed for
centralized learning models.

%--------------------------------------------------
\section{Conclusion}
%--------------------------------------------------
In this work, we present the first model hijacking attack against federated learning, namely \attack.
In this attack, an adversary (i.e., a malicious client) repurposes a global model designed for a specific task to perform an adversary-defined hijacking task.
The core design of \attack is to introduce pixel-level perturbations (cloaks) to hijacking samples, enabling their extraction and identification in a manner similar to the original samples within the feature space. Consequently, these samples are accurately classified into the corresponding original classes. We conduct extensive evaluations on four benchmark computer vision datasets and three widely used model architectures. Empirical findings demonstrate that our attack attains a high level of attack success rate without compromising model utility, meeting the two requirements for hijacking a federated learning model. Furthermore, we explore two possible defenses that can detect hijacking samples in the FL prediction phase.

Our goal with this work is to first raise awareness of the accountability and parasitic computing risks that can be posed by hijacking attacks in the federated learning community.
We further discuss the potential defense mechanism with the aim of motivating the community to investigate different mitigation techniques to address these risks.

%--------------------------------------------------
\bibliographystyle{plain}
\bibliography{sample-base}
%--------------------------------------------------

\begin{figure*}[!th]
    \centering
    \includegraphics[width=0.9\linewidth]{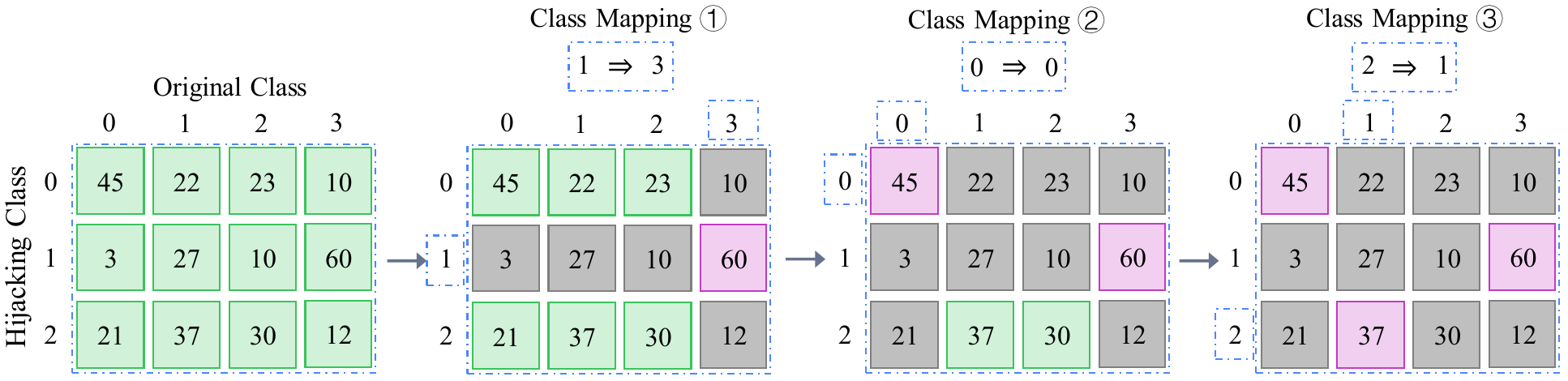}
    \caption{ Class Mapping Steps
    }
    \label{fig:relabel_details}
\end{figure*}

\newpage
\appendix

\begin{table}[!t]
    \centering
\caption{The threshold of the feature-based anomaly defense..}\label{tab:threshold_feature_defense}
\scalebox{0.7}
    {
\begin{tabular}{l|c|c}
\toprule
TaskSet             & ResNet-18      & MobileNet-V2        \\
\cmidrule(lr){1-3}
\RNum{1}            & 0.6                & 0.6       \\
\cmidrule(lr){1-3}
\RNum{2}                         &        0.6     & 0.6        \\
\cmidrule(lr){1-3}
\RNum{3}                        &      0.6      & 0.6       \\
\cmidrule(lr){1-3}
\RNum{4}                         &     0.6     & 0.6      \\
\bottomrule  
\end{tabular}
}
\end{table}

\begin{table}[!t]
    \centering
\caption{The threshold of Feature Squeezing.}\label{tab:thres_fs_defense}
\scalebox{0.7}
    {
\begin{tabular}{l|c|c|c|c}
\toprule
\multirow{2}{*}{TaskSet} & \multicolumn{2}{c|}{ResNet-18} & \multicolumn{2}{c}{MobileNet-V2} \\
\cmidrule(lr){2-5}
                         & Threshold 1      & Threshold 2      & Threshold 1     & Threshold 2        \\
\cmidrule(lr){1-5}
\RNum{1}                         & 3      & 15          &    3       & 15        \\
\cmidrule(lr){1-5}
\RNum{2}                         &     3    & 15       &      3       & 15        \\
\cmidrule(lr){1-5}
\RNum{3}                        &      20      & 60      &   20        & 60        \\
\cmidrule(lr){1-5}
\RNum{4}                         &     20     & 60      &       20      & 60     \\
\bottomrule  
\end{tabular}
}
\end{table}

\subsection{The Convergence of Cloak Optimization}\label{sec:convergence}
In this section, we delve into the convergence properties of cloak optimization. 
\autoref{eq_4} defines an optimization problem aimed at minimizing the weighted sum of distances between two features. 
Specifically, we employ the L2 distance as our loss function to quantify these distances.

\mypara{Loss Function Convexity}
The L2 loss function is a squared loss function.
Since the squared function is convex, the L2 loss function inherits this property. 
Convex functions exhibit desirable optimization properties, ensuring that the optimal solution is globally optimal and avoids local minima. 
This helps the optimization algorithm to converge to the global optimum.

\mypara{Gradient Continuity} 
This gradient is continuous because the derivative of the squared function is a linear function that can be derived everywhere.
The continuity of the gradient ensures that the gradient information is reliable in the process of finding the optimal solution of the optimization algorithm, and there will not be a situation where the gradient does not exist or is not continuous.

\mypara{Convergence} 
Since the L2 loss function is convex and the gradient is continuous, optimization algorithms based on gradient descent can converge to the global optimum.
The continuous gradient guides the optimization process toward the optimal solution along the descent direction until convergence. 
Importantly, convexity prevents the algorithm from getting trapped in local optima.

In summary, the L2 distance as a loss function has good mathematical properties, i.e., convexity and continuity of the gradient, which makes the optimization algorithm based on gradient descent able to converge to the global optimal solution. 
This is one of the important reasons why the L2 loss function is widely used in machine learning and optimization problems.
\end{document}